\documentclass{egpubl_new}
\usepackage{amsmath}
\usepackage[bb=px]{mathalfa}
\usepackage{multirow}

\usepackage[T1]{fontenc}
\usepackage{dfadobe}  

\usepackage{cite}  
\BibtexOrBiblatex
\electronicVersion
\PrintedOrElectronic
\ifpdf \usepackage[pdftex]{graphicx} \pdfcompresslevel=9
\else \usepackage[dvips]{graphicx} \fi

\usepackage{egweblnk} 

\title[Generative Adversarial Shaders for Real-Time Realism Enhancement]%
      {Generative Adversarial Shaders for\\Real-Time Realism Enhancement}

\author[A. Salmi, Sz. Cs\'efalvay \& J. Imber]
{\parbox{\textwidth}{\centering A. Salmi, Sz. Cs\'efalvay and J. Imber 
        }
        \\
{\parbox{\textwidth}{\centering Imagination Technologies, United Kingdom\\
                     \{Arturo.Salmi, Szabolcs.Csefalvay, James.Imber\}@imgtec.com}
}
}

\begin{document}

\teaser{
 \includegraphics[width=\linewidth]{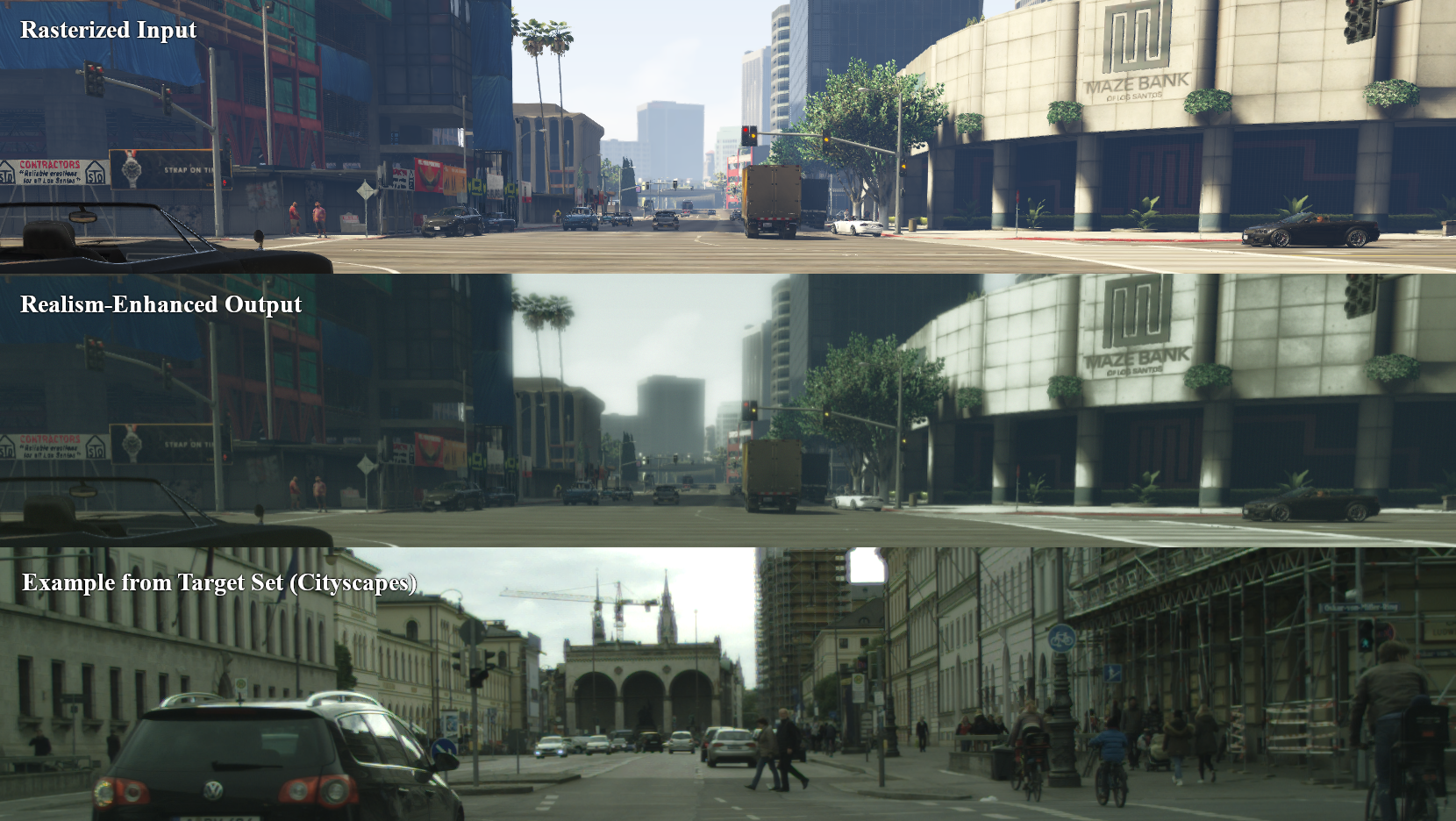}
 \centering
  \caption{Our proposed Generative Adversarial Shaders combine the quality and automation of an adversarial training methodology with the computational efficiency and performance of a real-time post-processing shader pipeline. This crop is from a 1080p frame which was processed in 2.8ms on an Imagination Technologies BXT-32-1024 MC4 GPU to match the style of the Cityscapes dataset \cite{Cityscapes}.}
\label{fig:teaser}
}

\maketitle
\begin{abstract}
Application of realism enhancement methods, particularly in real-time and resource-constrained settings, has been frustrated by the expense of existing methods. These achieve high quality results only at the cost of long runtimes and high bandwidth, memory, and power requirements. We present an efficient alternative: a high-performance, generative shader-based approach that adapts machine learning techniques to real-time applications, even in resource-constrained settings such as embedded and mobile GPUs. The proposed learnable shader pipeline comprises differentiable functions that can be trained in an end-to-end manner using an adversarial objective, allowing for faithful reproduction of the appearance of a target image set without manual tuning. The shader pipeline is optimized for highly efficient execution on the target device, providing temporally stable, faster-than-real time results with quality competitive with many neural network-based methods.


\end{abstract}  
\section{Introduction} 

Achieving greater photorealism has been a core concern of computer graphics research almost since the inception of the field. However, despite significant advances due to factors such as improved hardware capabilities, rendering algorithms, material representations and artistic tooling, there remains a stubborn shortfall in photorealism in real-time applications. This is particularly pronounced in settings with limited resources (e.g. bandwidth, compute, and power), such as smartphone GPUs.

There is a long history of using post-processing techniques to enhance the perceived realism of renders, for example by using programmable shaders to add such effects as chromatic aberration, lens flare, film grain, bokeh and motion blur to rendered images \cite{McIntosh2012} \cite{Guertin2014}. It is notable that many such techniques mimic artifacts of the image capture process rather than natural appearance. These methods are designed and tuned by hand in what can be a labor-intensive process, and the final result is often as much a question of artistic preference as mimicking a physical camera.

A compelling alternative approach to post-processed realism enhancement is afforded by conditional generative models. Richter et al. \cite{Richter_2021} demonstrate that image-to-image translation techniques utilizing neural networks and adversarial training \cite{IsolaTtoT} can bridge the realism gap without the introduction of artifacts present in comparable methods \cite{Park_2019_CVPR}. Furthermore, the appearance of the final renders can be matched to that of a target dataset. Their method achieves an impressively high degree of realism and visual quality, but would be prohibitively computationally expensive for real-time deployment on datacenter, desktop and embedded systems, particularly those without AI accelerators.

We hypothesize that much of the improvement in realism achieved by such deep learning-based methods may be accounted for by means of a carefully engineered shader pipeline exploiting domain knowledge, in particular by modelling the image capture process. A key advantage of conditional GANs \cite{park2020cut} \cite{jiang2020tsit} \cite{Richter_2021} over such a solution is that the manual tuning process is replaced with numerical optimization. This suggests a hybrid approach that combines the computational efficiency of a bespoke post-processing shader pipeline with the automation and quality of an adversarial training methodology. We also take inspiration from the physical capture models \cite{Hansen2021} that are occasionally used in the tuning of image signal processors.

With the objective of preserving the learning capabilities of neural networks while designing a model that can be used for real-time rendering, we therefore propose a method based on a learnable post-processing pipeline composed of lightweight functions with a limited parameter set that we term Generative Adversarial Shaders (GAS) (Figure \ref{fig:teaser}). By designing these functions to be differentiable, the pipeline can be trained in the same way as a neural network . We show that our method can be trained in an adversarial framework to learn a mapping between renders and photographic images. This provides several advantages over more conventional deep generative models: 

 \begin{itemize}
\item Improved computational efficiency over an undifferentiated neural network. Application of domain knowledge reduces the number of parameters by several orders of magnitude. The number of operations is brought well within the range that can be processed in real time by an embedded GPU alongside a rendering workload.
\item High degree of interpretability, debuggability and explainability due to its modular structure and low parameter count. Each stage in the pipeline has a clearly defined task by virtue of the function it performs. Parameters are typically semantically meaningful.
\item Reduced scope for overfitting and introduction of artifacts by virtue of the limited functional range of the solution. The flip side of this is that we are not able to model more complex effects such as volumetric foliage. We suggest that these can be added by inclusion of a small auxiliary neural network operating alongside the proposed method, but this is left to future work.
\end{itemize}

\begin{figure}[b]
  \centering
  \includegraphics[width=1\linewidth]{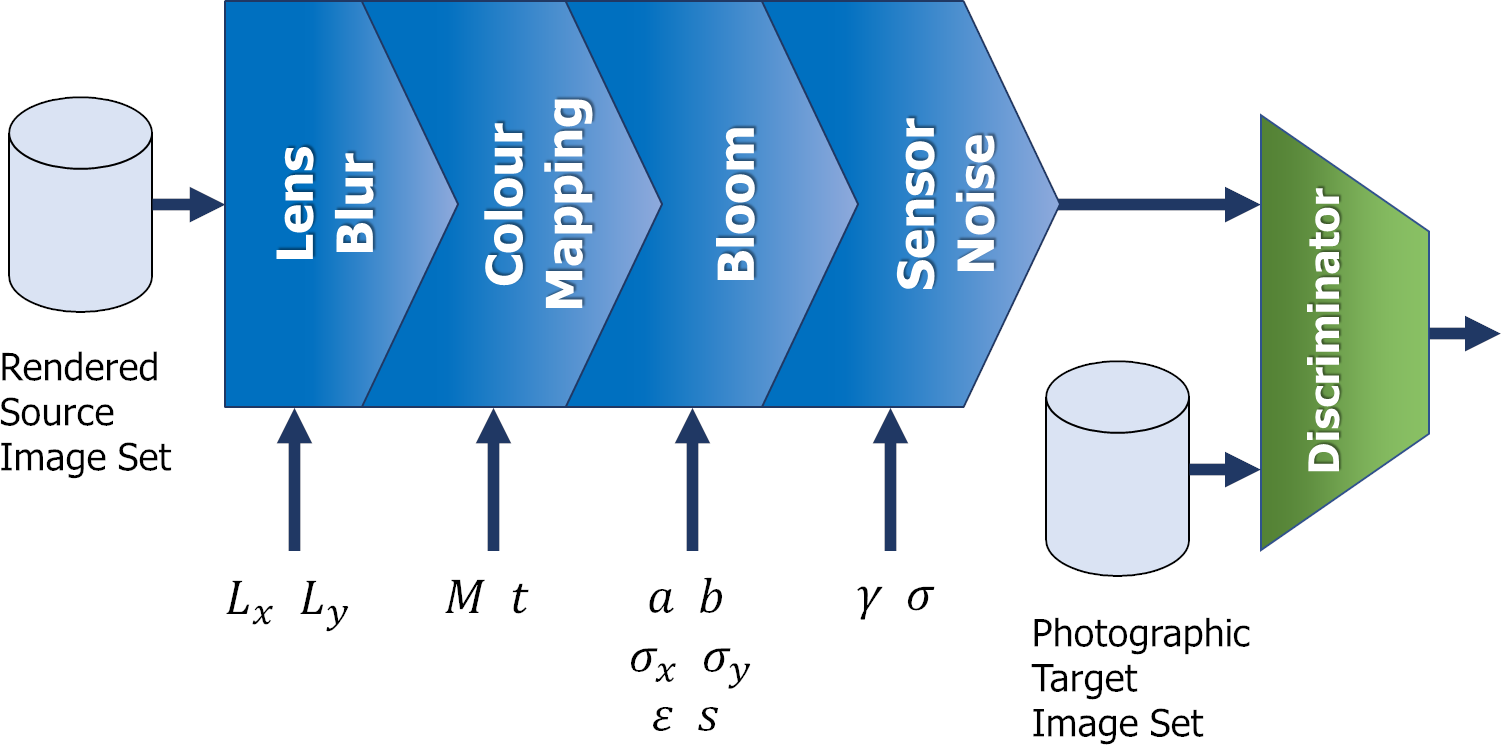}
  \caption{
    In our approach, the proposed differentiable post-processing pipeline (blue) comprising Generative Adversarial Shaders is trained as a generator in a conditional GAN. After training, the pipeline with learned parameters is deployed in a performance-optimized form (in our case, as OpenCL kernels).
  }
  \label{fig:shader_pipeline}
\end{figure}

The resulting shader pipeline (Figure \ref{fig:shader_pipeline}) can then, for example, be deployed in a conventional manner, for example in OpenCL on a GPU. We only require final renders as inputs, which improves the ease of integration of our method. We do not require metadata such as G-buffer renders. The results we obtain are a step in the direction of real-time photorealism on resource-constrained platforms. 

\section{Related Work}

Physically-based methods are a widespread approach for synthesizing photorealistic images. These simulate the physical processes that are involved in image formation, such as light transport \cite{DBBAGIbook2ndEdition}, \cite{Myszkowski2001b} and other physical effects such as translucent media \cite{Harris2001}.  The quality of the images synthesized by these methods depends on the approximation of the physical models used:  accurate modelling will produce realistic results, but these also typically come with large costs in terms of computational requirements and technical and artistic labor. Approximations of these methods can obtain real-time or interactive-time methods with an acceptable level of realism \cite{ReinhardRealsim}.

Conditional image synthesis methods (for example, conditional GANs) can produce photorealistic images given a corresponding input image in a different domain, commonly referred to as \textit{image-to-image translation} \cite{IsolaTtoT} \cite{park2020cut} \cite{jiang2020tsit} \cite{huang2018munit}. Recent advances in diffusion models and attention-based architectures for conditional image generation have also been applied to the image-to-image translation problem \cite{Torbunov2023} \cite{Parmar2023}. Such image-to-image methods are applicable to post-processed realism enhancement of rendered images \cite{Hoffman_cycada2017}. Park et al. \cite{Park_2019_CVPR} extend the idea to reliable generation of realistic synthetic images conditioned on semantic label maps; however when applied to the realism enhancement problem this leaves the problem largely unconstrained, reducing artistic control and in some cases impacting quality. 

Despite giving plausible results in many cases, deep image-to-image models are prone to jarring artifacts such as hallucinations which would be intolerable for many applications (noted by Richter et al. \cite{Richter_2021}). By using metadata extracted from the G-buffers to provide more scene information to the generator and further constrain the image formation process, Richter et al. overcome these recurrent issues and achieve impressive results both in terms of spatial and temporal stability and level of realism achieved. However, the proposed neural network is too expensive in terms of computational requirements and execution time to be feasible for real-time rendering applications.

Other techniques apply a data-driven approach to style transfer and realism enhancement. Early work showed that matching color distributions of renders to photographic examples can enhance the realism of computer graphics images \cite{ReinhardColour}, \cite{Piti2005NdimensionalPD}, \cite{PITIE2007123}.  With the adoption of CNNs, more sophisticated techniques for style transfer have been proposed \cite{Gatys_2016_CVPR}, \cite{Huang2017ArbitraryST}, \cite{SEAN}. Recent work has also focused on adopting an encoder/decoder structure to transfer the style of a target image while maintaining the content of the source image \cite{Yijun_2018}, \cite{Yoo_2019_ICCV}. Li et al. \cite{li2018learning} propose a linear transformation method, where the mapping is locally modulated by a controller CNN. These approaches often depend upon a favorable overlap between the source and the target images. Large differences in content or layout can be detrimental for the quality of the results produced by these methods. This makes these methods less suitable for dynamic scene rendering and application to a large variety of rendered scenes. 

Our proposed method combines traditional post-processing techniques with an adversarial training approach. In common with Carlson et. al. \cite{Carlson_2019} we identify modelling of the image capture process with an appropriate series of post-processing stages as being a key component of realism enhancement. However, we go further in proposing that our pipeline be differentiable, with the parameters directly optimized to reduce an adversarially defined loss. Nalbach et al. \cite{Nalbach2017} proposed a learnable post-processing method where the entire post-processing pipeline is approximated using a "Deep Shader" neural network. Instead, we design each shader as a bespoke learnable function, so that the pipeline can be trained in the same way as a neural network while retaining explainability and computing performance. Work from Thomas et al. \cite{Thomas2017} approximates global illumination using conditional generative adersarial networks with the goal of achieving offline rendering quality at interactive rates. While both adopt adversarial training, our method differs from Thomas et al. approach as it is designed to replicate post-processing effects applied in screen-space.

\begin{figure*}[h!]
  \begin{tabular}{p{0.3\textwidth}p{0.3\textwidth}p{0.3\textwidth}}
  \centering\textbf{Rasterized Input} & \centering\textbf{Lens Blur Output} & \centering\textbf{Example from target set Cityscapes}
  \end{tabular}
  \centering
  \includegraphics[width=\linewidth]{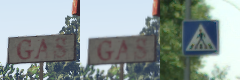}
  \caption{\label{fig:lens_blur}%
           Example of the learned lens blur applied to a rasterized frame from the GTA5 dataset \cite{GTA5}. The jagged edges of geometry are mitigated by the learned kernel. The foliage in the background also looks more similar to the foliage in the target dataset. 
           }
\end{figure*}

\section{Method}

Our proposed method has a learnable post-processing pipeline at its core (Figure \ref{fig:shader_pipeline}). Each shader in the pipeline is designed to replicate an aspect of the physical image capture process. The behavior of each shader is governed by a corresponding set of parameters which, due to the differentiability of the pipeline, can be optimized during training in a conventional manner using backpropagation of errors. We replace the generator in a conventional conditional GAN with this pipeline. In this way, each shader is trained to optimally replicate the relevant aspect of the target dataset.

The shader pipeline is implemented in a deep learning framework such as PyTorch. However, we note that while this implementation is convenient for training, this is not optimal for performance, particularly on a resource-constrained target platform. For this reason, we also implement the shaders as functionally-equivalent OpenCL compute kernels optimized for inference speed. The stages in the pipeline can be merged at deployment to prevent excess system bandwidth consumption. After training, the learned parameters are applied to this deployed version of the shader pipeline. Our approach is robust against strong artifacts and hallucinations frequently observed in deep learning-based approaches, by virtue of our constraining the operation of each shader to a tightly-defined mathematical function. To further improve training stability, we also adopt regularization strategies which are described in more detail in section \ref{sub:training}.

We fully expect the proposed training and deployment methodology to be extensible beyond the functionality modelled by the pipeline detailed below, as described in section \ref{sub:limitation}.

\subsection{Learnable Shader Pipeline}
\label{sub:learnable_shader_pipeline}
Our shader pipeline consists of four stages, each taking as inputs learnable parameters which can be trained through backpropagation when implemented in a deep learning framework like PyTorch or TensorFlow. We fully expect the proposed pipeline to be extensible with more shaders implementing additional functionality (e.g. lens distortion, chromatic aberration, etc). The shaders are applied in approximately the order in which the corresponding physical process in image capture occurs.

\subsubsection{Lens Blur}

The optics of physical camera lenses introduce a certain amount of blur. This is modelled with a lens blur shader as the first module of the pipeline, which we implement as a learnable separable convolution (Equation \ref{eq:lens_blur}), where both kernels are initialized as unit impulses to avoid introducing translation at the start of training. The window size is set to 5 since this was found to give a good balance between quality and performance in the deployed system.

\begin{equation}
\label{eq:lens_blur}
I_{out} = \left(I_{in} \ast L_{x}\right) \ast L_{y}
\end{equation}

In Equation \ref{eq:lens_blur}, ${I_{in}}$ and ${I_{out}}$ are the input and output images, ${L_x}$ and ${L_y}$ are the horizontal and vertical blur kernels (represented with 5 values each), and ${*}$ is the convolution operation. During training, all weights of both kernels are updated. The weights of each kernel are normalized to sum to one before convolution. This is done to maintain output image intensity, as the shader should not be changing the overall brightness of the input frame. As well as replicating the blurring artifacts introduced by real world camera lenses, the learned convolutional operator also acts as a cheap anti-aliasing filter for point-sampled geometry. As we show in Figure \ref{fig:lens_blur}, this shader can learn to convolve the image in order to mitigate jagged rendering artifacts at geometry edges in the input image.

\subsubsection{Color Mapping}

A post-processing pipeline for realism enhancement should transfer the color characteristics of photographic examples to the input frames. Rather than operating on a single source and target example \cite{ReinhardColour}, our color mapping shader learns a fixed transformation of the color distribution from the source to the target dataset. We learn a full affine mapping capable of modelling brightness and contrast adjustment, to avoid restricting to purely linear \cite{Hansen2021} \cite{ReinhardColour} or translational \cite{Carlson_2019} mappings. Our proposed affine transformation (Equation \ref{eq:affine_colour_model}), comprises a 3x3 matrix for the linear part of the mapping, and a 3-element vector for the translational part. All 12 parameters of the model are learnable.

\begin{equation}
\label{eq:affine_colour_model}
\begin{bmatrix}
R^{'}_{x,y} \\
G^{'}_{x,y} \\ 
B^{'}_{x,y}
\end{bmatrix}
=
\begin{bmatrix}
m_{11} && m_{12} && m_{13} \\
m_{21} && m_{22} && m_{23} \\
m_{31} && m_{32} && m_{33} 
\end{bmatrix}
\begin{bmatrix}
R_{x,y} \\
G_{x,y} \\ 
B_{x,y}
\end{bmatrix}
+
\begin{bmatrix}
t_{1} \\
t_{2} \\ 
t_{3}
\end{bmatrix}
\end{equation}

\begin{figure*}[h!]
  \centering
  \includegraphics[width=\linewidth]{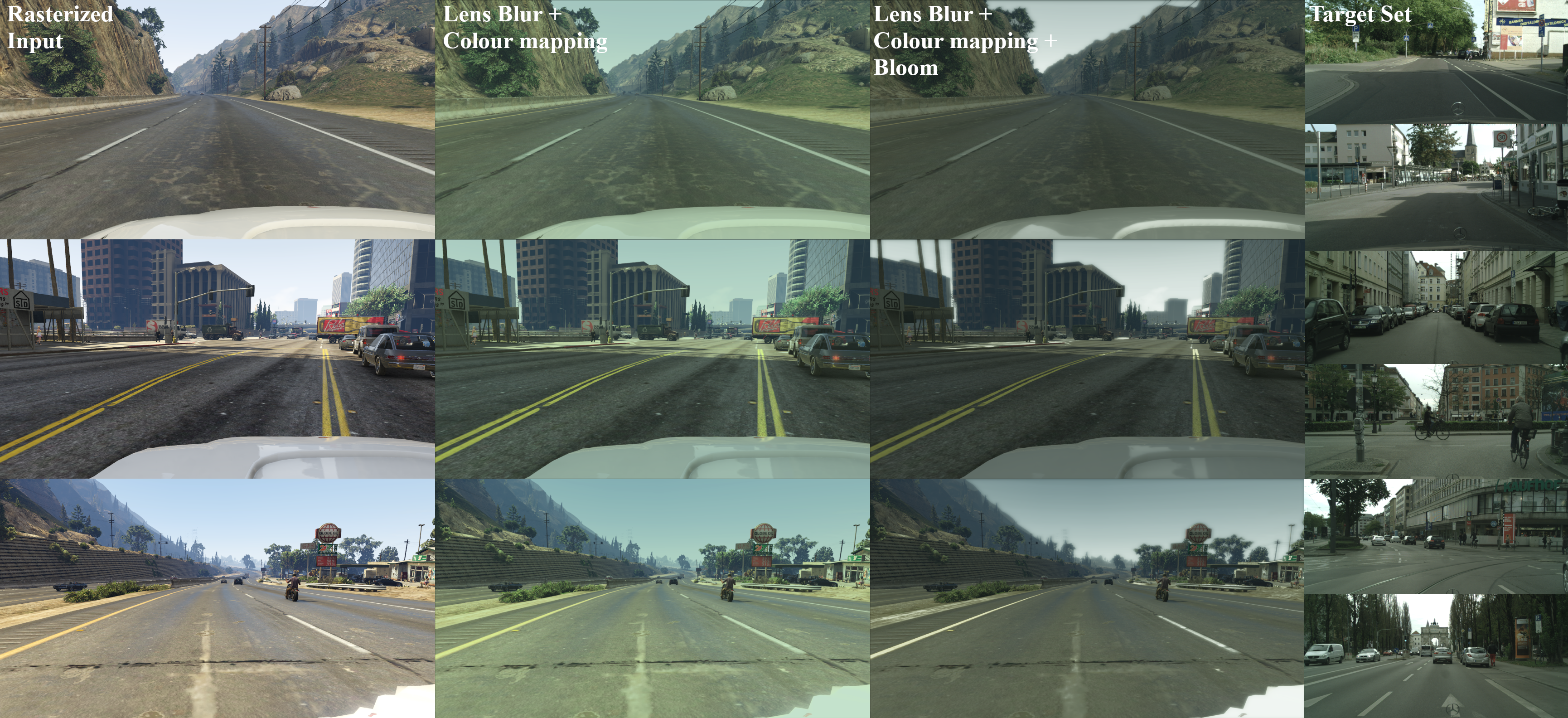}
  \caption{\label{fig:colour_map_bloom_comparison}%
           The color mapping function transfers the global color characteristics of the 
           target dataset to the rasterized input frames. As well as replicating the "bleeding"
           effect caused by overexposure, by separately processing bright regions, the bloom
           shader helps prevent them from distorting the learned color mapping. 
           }
\end{figure*}

Since this operates in a pixel-wise manner without reference to the local neighborhood, it has no impact on the general structure of the image. We initialize the transform with the identity function to aid convergence. Some examples of our learned color mappings are shown in Figure \ref{fig:colour_map_bloom_comparison}. Best results were obtained when the color mapping shader was trained in conjunction with the bloom shader.

\subsubsection{Bloom}

Our bloom shader approximates the behavior of cameras when capturing scenes with bright, saturated light sources. The bloom effect is commonly implemented as a post-processing stage to enhance perceived realism in graphics \cite{BloomGPUGems}. In captured images it can often be seen that saturated regions "bleed" over to neighboring pixels, causing a characteristic glowing effect. Our proposed bloom shader performs the following steps:

\begin{enumerate}
\item Isolate high intensity pixels (the glow sources) in the input image $I_{in}$ by soft thresholding with a sigmoid (Equation \ref{eq:glow_mask}) to obtain a "glow mask" $I_{glow}$.
\begin{equation}
\label{eq:glow_mask}
I_{glow} = I_{in} \cdot \left( \frac{1}{1 + e^{-b\left(I_{luma} - a\right)}} \right)
\end{equation}
In common rendering pipelines, $I_{glow}$ can be obtained by setting a minimum HDR value threshold, with pixels lower than this threshold set to black. The render dataset we used for training, however, only provides 8-bit tone-mapped SDR frames. Furthermore, we require that the threshold be learnable. To solve this issue, we use a sigmoid as the thresholding function applied to a luma channel $I_{luma}$  from the input frame $I_{in}$. $a$ and $b$ are learnable parameters.
\item Apply a parameterized blur using a separable convolution kernel then blend with the input image (Equation \ref{eq:bloom_mask}) to obtain $I_{bloom}$.
\begin{equation}
\label{eq:bloom_mask}
I_{bloom} = I_{in} + \left(I_{glow} * G_{x}\left(\sigma_x^2\right)\right) * G_{y}\left(\sigma_y^2\right)
\end{equation}
$G_x(\sigma_x^2)$ and $G_y(\sigma_x^2)$ are discrete 1D Gaussian kernels normalized to sum to one. The variances $\sigma_x^2$ and $\sigma_y^2$ of the kernels are learned separately. By doing so, the shader can learn different amounts of "glow" to apply horizontally and vertically.
\item Apply a non-linear colour mapping to $I_{bloom}$ (Equation \ref{eq:bloom_out}), obtaining $I_{out}$ with the finished bloom effect applied.
\begin{equation}
\label{eq:bloom_out}
I_{out} = min \left(\left(\frac{e^{\epsilon s}}{e^{\epsilon s} - 1}\right)\left(1 - e^{-I_{bloom}\epsilon}\right), 1\right)
\end{equation}
Exposure $\epsilon$ is a learnable parameter that controls the steepness of the curve and $s$ is a learnable saturation parameter that controls the saturation point of the curve.
\end{enumerate}

Since bloom can have a large radius of influence, for efficiency we use a multi-resolution approach for generating the bloom map in steps 1 and 2 above, before upsampling to full resolution and summing to produce a single bloom map for blending in step 3. For this purpose, 4 bloom maps were generated: one each at full, half, quarter and one-eighth resolution. Best results are achieved by learning separate sets of parameters for each resolution of the bloom algorithm.

\subsubsection{Sensor Noise}

Camera sensors introduce noise during image capture. A commonly-used noise model \cite{Carlson_2019} \cite{Hansen2021} breaks this into two components: (1) pixel capture noise, which is a Poisson process $P$ depending on intensity; and (2) normally distributed noise as a catch-all for other processes (such as read noise \cite{JanesickCCD} and thermal noise). The division by gain $\gamma$ in the Poisson distribution in our model (Equation \ref{eq:noise_model}) is to correct for intensity scaling between input $I_{in}$ and output $I_{out}$ due to gain.

\begin{equation}
\label{eq:noise_model}
I_{out} = \gamma \left( P\left(\frac{I_{in}}{\gamma} \right) + \mathcal{N}\left(0, \sigma'^2\right) \right)
\end{equation}

For improved performance and ease of implementation, we approximate the Poisson-distributed noise with a zero-mean normal distribution. We also reparameterize the standard deviation of the second term as $\sigma=\gamma\;\sigma'$ to remove the dependence on gain, which helps orthogonalise the model.

\begin{equation}
\label{eq:noise_change_of_variable}
I_{out} = I_{in} + \mathcal{N}\left(0, \gamma \; I_{in}\right) + \mathcal{N}\left(0, \sigma^2\right)
\end{equation}

Correct modelling of the noise, particularly readout noise in dark regions, depends on the input image $I_{in}$ being in linear color space. For this reason, we opt to apply inverse gamma before this shader, and reapply the gamma afterwards. The gain $\gamma$ and standard deviation $\sigma$ are the learnable parameters of this shader. To obtain valid gradients, we rearrange equation \ref{eq:noise_change_of_variable} to yield equation \ref{eq:noise_final}:

\begin{equation}
\label{eq:noise_final}
I_{out} = I_{in} + \sqrt{\gamma \; I_{in}} \mathcal{N}\left(0, 1\right) + \sigma \; \mathcal{N}\left(0, 1\right)
\end{equation}

%

\begin{table*}[t]
\centering
\begin{tabular}{c c || c | c | c}
 \hline
  & Method & KID score (x1000) & FID score & Mean execution time \\ [0.5ex] 
 \hline
  & GTA \cite{GTA5} & $76.88 \pm 0.11$ & 71.67 & n/a \\ 
 \hline
  \multirow{2}{*}{Color transfer} &  Color Transfer \cite{ReinhardColour} & $70.08 \pm 0.15$ & 66.12 & 0.183ms $\pm$ 0.042ms \\
  &  CDT \cite{PITIE2007123} & $85.98 \pm 0.08$ & 76.91 & 1003ms $\pm$ 3.9ms \\[1ex]
  Photo style transfer &  WCT2 \cite{Yoo_2019_ICCV} & $62.71 \pm 0.08$ & 64.50 & 1650ms $\pm$ 5ms \\[1ex]
  \multirow{5}{*}{Image-to-Image Translation} & CyCADA \cite{Hoffman_cycada2017} & $27.98 \pm 0.09$ & 33.47 & 29.12ms $\pm$ 0.5ms \\
  &  CUT \cite{park2020cut} & $36.68 \pm 0.10$ & 41.25 & 98.9ms $\pm$ 3ms \\ 
  &  MUNIT \cite{huang2018munit} & $33.87 \pm 0.07$ & 41.87 & 167ms $\pm$ 5ms \\
  &  TSIT \cite{jiang2020tsit} & $46.29 \pm 0.10$ & 48.80 & 227ms $\pm$ 2ms \\
  &  EPE $\dag$ \cite{Richter_2021} & $31.57 \pm 0.08$ & 37.73 & 1268ms \\
 \hline
  &  \textbf{Ours} & $59.67 \pm 0.09$ & 61.65 & \textbf{0.089ms} $\pm$ \textbf{0.0013ms} \\ 
 \hline
\end{tabular}
\caption{KID and FID scores measured for domain adaptation methods using Cityscapes \cite{Cityscapes} as the target distribution. Execution times were measured on an Nvidia RTX 2080 GPU at 960x540 resolution, on which our proposed method (designed for 1080p on mobile GPUs) achieves well in excess of real-time performance. $\dag$ These KID and FID scores were obtained on JPEG-compressed images, and execution time was estimated based on scaling in FLOPs from that reported in \cite{Richter_2021}.}
\label{tab:KID}
\end{table*}

\subsection{Training}
\label{sub:training}

The learnable shader pipeline (Section \ref{sub:learnable_shader_pipeline}) acts as the generator in an adversarial learning framework. Since care has been taken to ensure that the pipeline is differentiable, the parameters of the pipeline are updated in a conventional manner using backpropagation of errors. Using the information provided by the adversarial loss, the pipeline parameters are optimized so that they minimize the difference between the post-processed and the real image sets. PatchGAN \cite{LiW16},\cite{IsolaTtoT} was chosen as the discriminator, as by penalizing the pipeline on patches we can improve the local quality of effects such as bloom, as well as global details like color matching.

There is a gap in capacity and functional flexibility between the shader pipeline and the discriminator. This can lead to training instability, as the discriminator has the potential to exploit inflexibility in the generator, making its task of distinguishing real from post-processed images easier. To limit this behavior, we added instance noise to our discriminator \cite{sonderby2014apparent},\cite{arjovsky2017towards},\cite{Mescheder2018ICML}, with the only difference that rather than adding noise to the logits of the discriminator we add noise between the layers. This has the effect of smoothing the distributions of the source and target datasets, making them more similar and hence harder to distinguish.

The images used during training are 256x256 random crops from the input data frames. This helps prevent the discriminator from memorizing the dataset. Before being fed into the discriminator, the edges of the pipeline output are cropped to prevent the discriminator from learning how to distinguish generated and target images based by any edge effects present in the target or post-processed image sets. Spectral normalization was used to stabilize the discriminator \cite{miyato2018spectral}.

For the training loss, we adopted a relativistic discriminator loss inspired by the relativistic least-square GAN loss used by Jiang et al. \cite{jiang2021enlightengan} (following LSGAN \cite{LSGAN2017}), as this was shown to perform well at matching local characteristics of the generated images in unpaired training settings.

The pipeline and the discriminator are trained using the \textit{ADAM} optimizer with learning rate set to $10^{-4}$ for both models. The training is continued until convergence of the adversarial loss.



\section{Results}


We evaluate the performance of our model on the unsupervised domain adaptation task (specifically rasterized to photographic images), and compare it with other methods proposed to solve the same task. Specifically, we evaluate our method on mapping from the \textit{GTA5} dataset \cite{GTA5} to the \textit{Cityscapes} dataset \cite{Cityscapes}.

\subsection{Quantitative Evaluation}

FID \cite{NIPS2017_8a1d6947} and KID \cite{binkowski2018demystifying} are two commonly-used metrics used to evaluate the performance of generative models. Recent work by Parmar et al. however showed that these scores can vary significantly depending on factors such as the type of pre-processing applied to the images used for evaluation and/or the compression scheme used on the image \cite{parmar2021cleanfid}. For reasons of reproducibility and stability, we used the evaluation methodology and code provided by Parmar et al., both for FID and KID, without modification. Furthermore, wherever possible the evaluation is conducted on images stored in lossless formats in order to avoid any score variation introduced by compression schemes. 

%

\begin{table*}[h!]
\centering
\setlength\tabcolsep{0pt}
\begin{tabular}{c c c c c c}
 \hline
  & \textbf{Rasterized Input} & \textbf{Lens Blur} & \textbf{Color Mapping} & \textbf{Bloom} & \textbf{Noise} \\ [0.5ex] 
  & \includegraphics[width=.19\linewidth]{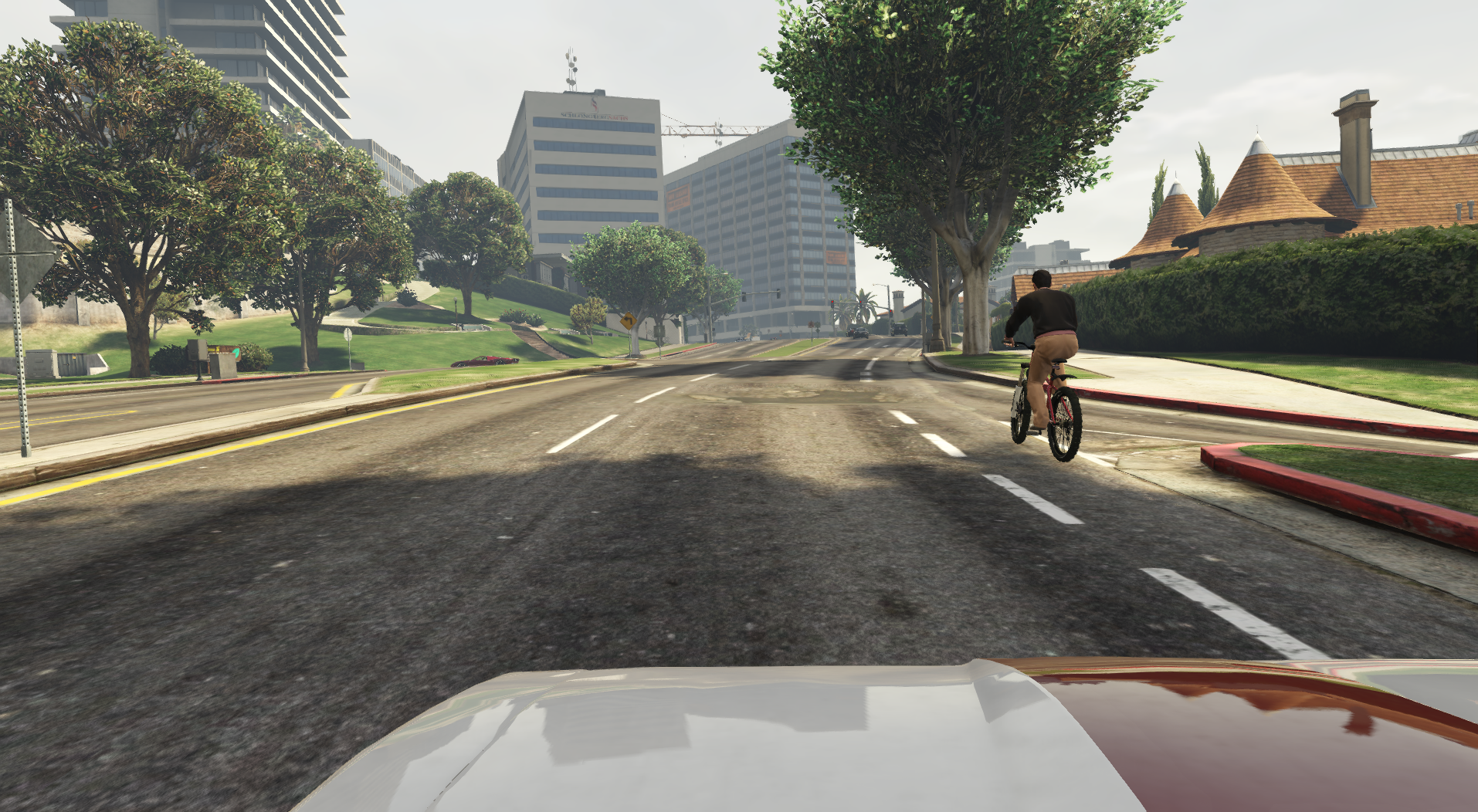} & \includegraphics[width=.19\linewidth]{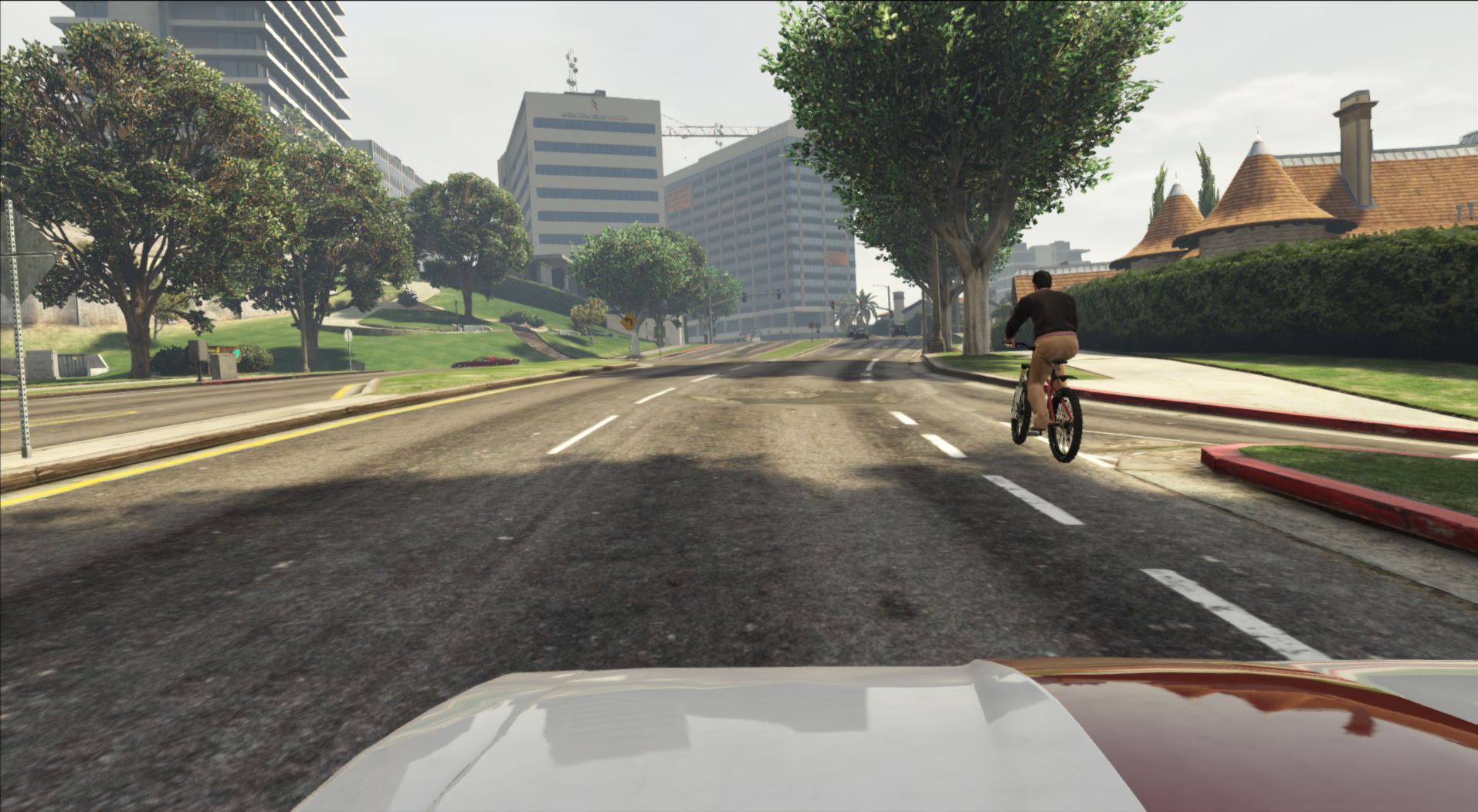} & \includegraphics[width=.19\linewidth]{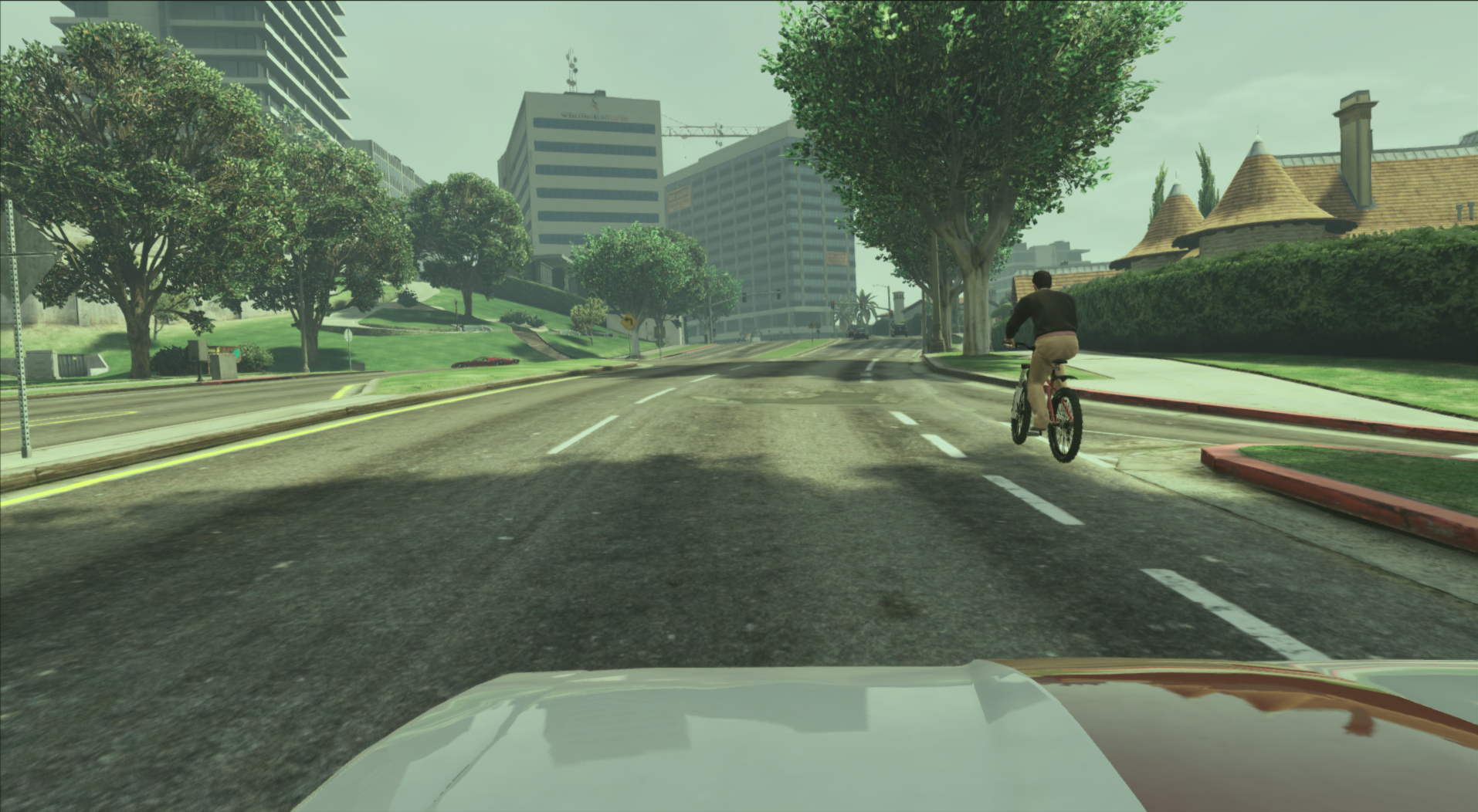} & \includegraphics[width=.19\linewidth]{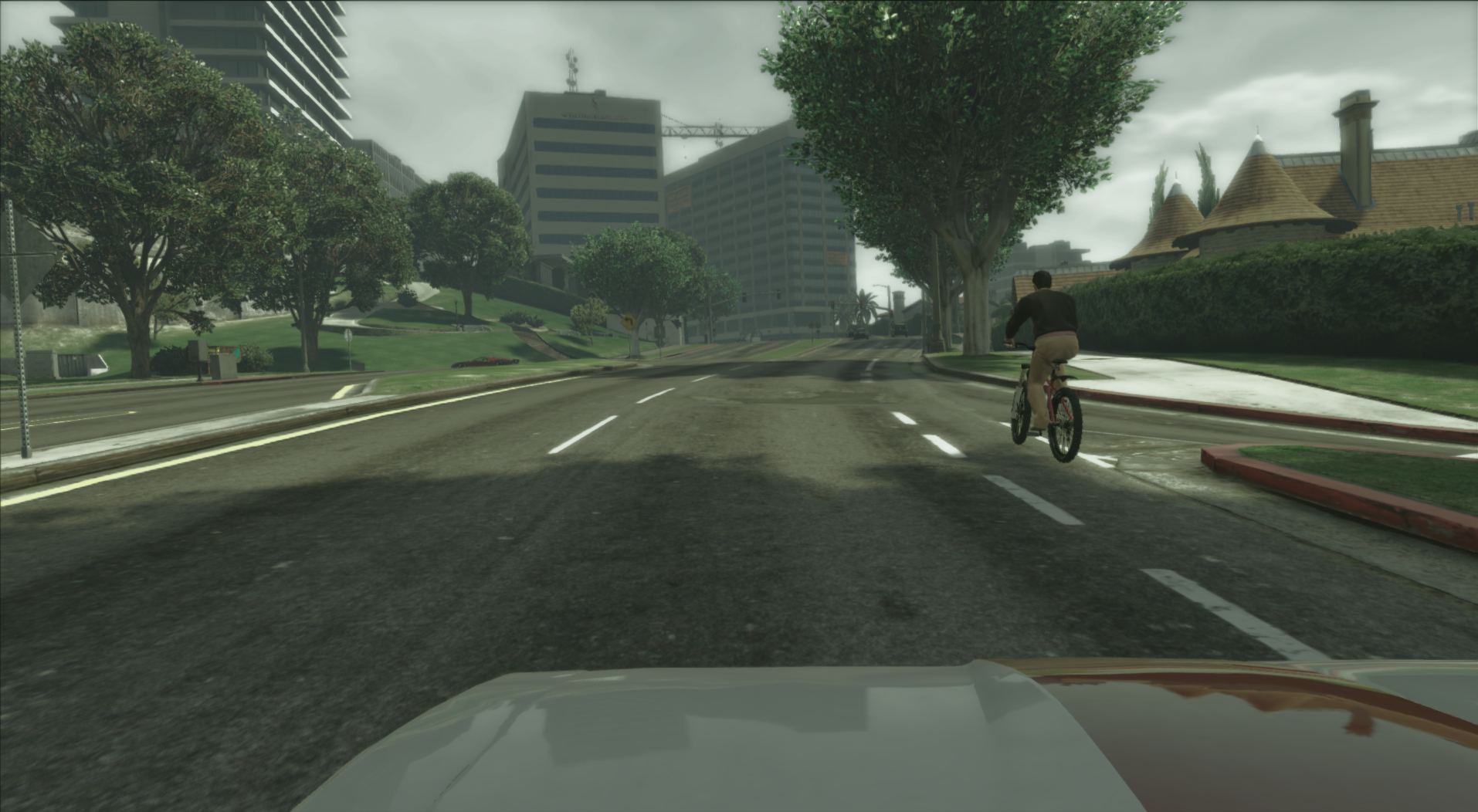} & \includegraphics[width=.19\linewidth]{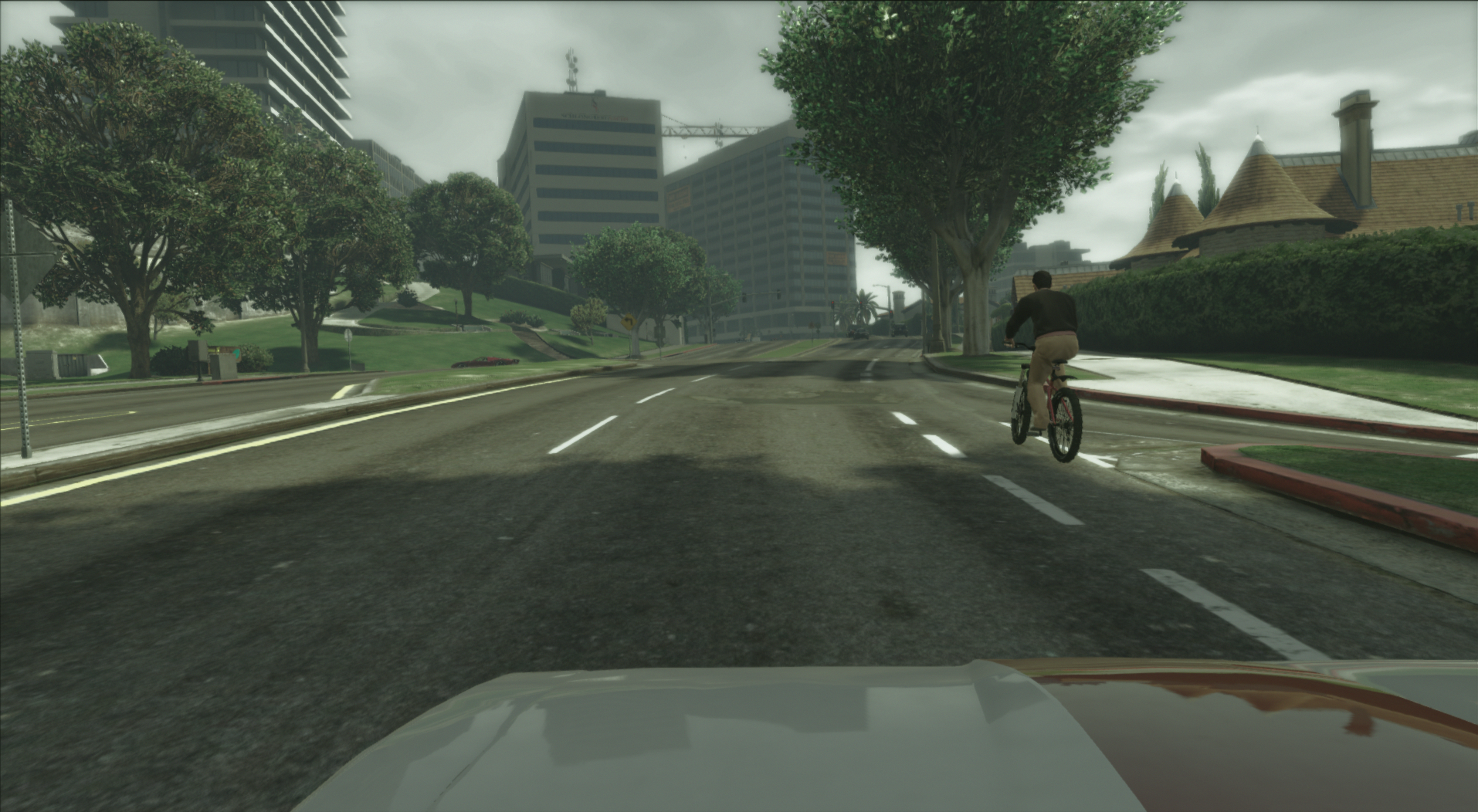}\\
  & \includegraphics[width=.19\linewidth]{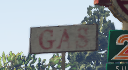} & \includegraphics[width=.19\linewidth]{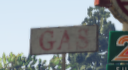} & \includegraphics[width=.19\linewidth]{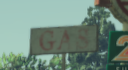} & \includegraphics[width=.19\linewidth]{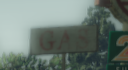} & \includegraphics[width=.19\linewidth]{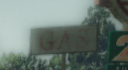}\\
  & \includegraphics[width=.19\linewidth]{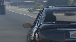} & \includegraphics[width=.19\linewidth]{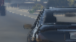} & \includegraphics[width=.19\linewidth]{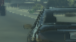} & \includegraphics[width=.19\linewidth]{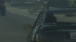} & \includegraphics[width=.19\linewidth]{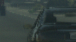}\\
  & \includegraphics[width=.19\linewidth]{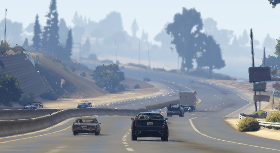} & \includegraphics[width=.19\linewidth]{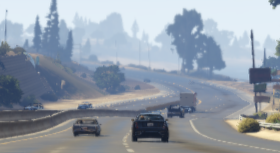} & \includegraphics[width=.19\linewidth]{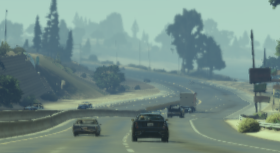} & \includegraphics[width=.19\linewidth]{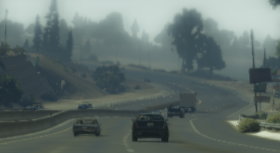} & \includegraphics[width=.19\linewidth]{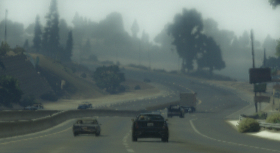}\\[0.5ex] 
  \hline
  \textbf{KID} & $76.88 \pm 0.11$ & $76.18 \pm 0.12$ & $67.31 \pm 0.08$ & $58.46 \pm 0.09$ & $59.67 \pm 0.09$ \\
  \textbf{FID} & $71.67$ & $71.55$ & $65.39$ & $59.05$ & $61.65$ \\
 \hline
\end{tabular}
\caption{Ablation study of the proposed pipeline. Shaders are added one at a time from left to right, and the pipeline trained from scratch each time. 
         The KID and FID score is measured on Cityscapes for the resulting enhanced images to analyze each shader's contribution. The resulting scores
         are below the images.}
\label{tab:ablation}
\end{table*}

\begin{figure}[h]
  \centering
  \includegraphics[width=1\linewidth]{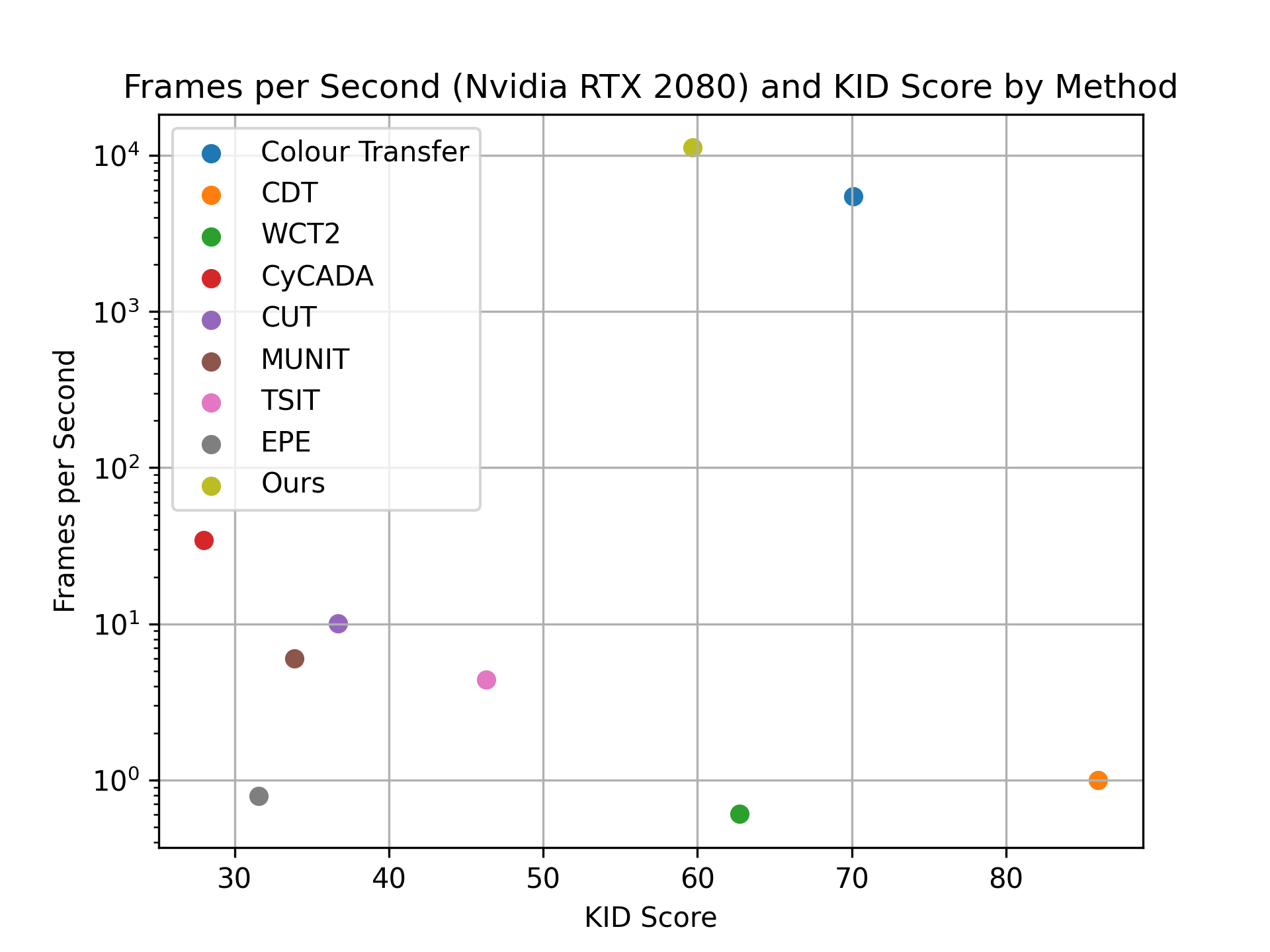}
  \caption{
    A graphical presentation of the data in Table \ref{tab:KID}. Note the logarithmic y-axis.
  }
  \label{fig:plot_score}
\end{figure}

Table \ref{tab:KID} contains the scores and execution times of each method. The proposed method is faster than other methods in the evaluation, and in particular is orders of magnitude faster than deep-learning based methods (Figure \ref{fig:plot_score}). In addition to its fast execution time, it also achieves a respectable KID and FID score. We note, however, that KID and FID are forgiving of many unacceptable artifacts present in some of the deep-learning based methods, such as hallucinated scenery. Some deep learning models will introduce extraneous objects in the scene (such as trees in place of sky) which do not damage the measured score. These scores are also not indicative of temporal stability, where our algorithm excels.

\begin{figure*}[h!]
  \begin{tabular}{p{0.3\textwidth}p{0.3\textwidth}p{0.3\textwidth}}
  \centering\textbf{Rasterized Input} & \centering\textbf{Ours} & \centering\textbf{EPE\cite{Richter_2021}}
  \end{tabular}
  \centering
  \includegraphics[width=\linewidth]{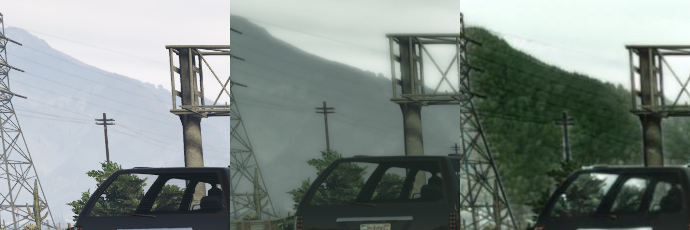}
  \caption{\label{fig:halo_comparison}%
           EPE produces halo artifacts around certain objects in the input frame.
           This is likely to be caused by the segmentation maps used by their proposed generator.
           }
\end{figure*}

\subsection{Qualitative Evaluation}

\begin{figure*}[tbp]
  \centering
  \includegraphics[width=.37\linewidth]{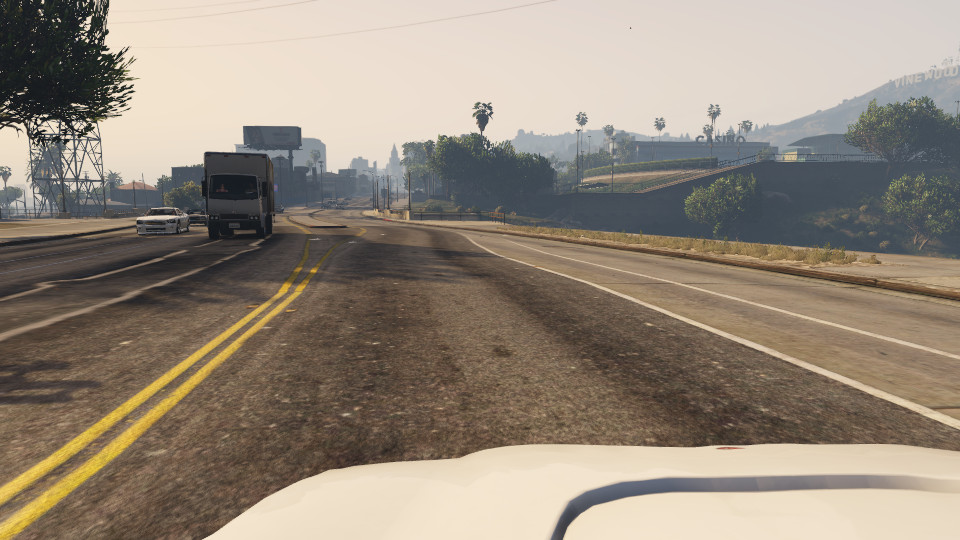} \hspace{5.5em}
  \includegraphics[width=.37\linewidth]{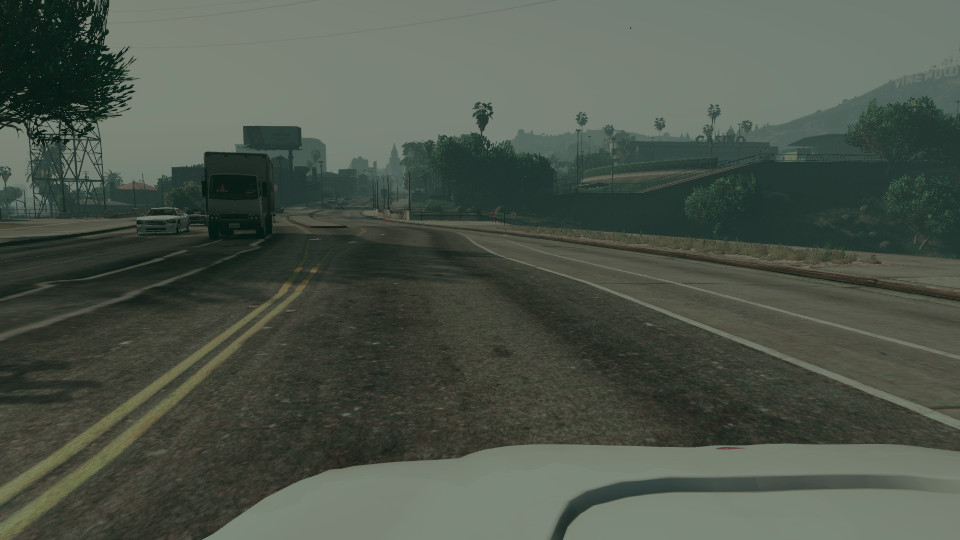} \\
  \begin{tabular}{p{0.45\textwidth}p{0.45\textwidth}}
  \centering GTA \cite{GTA5} & \centering Colour Transfer\cite{ReinhardColour}
  \end{tabular}
  \includegraphics[width=.37\linewidth]{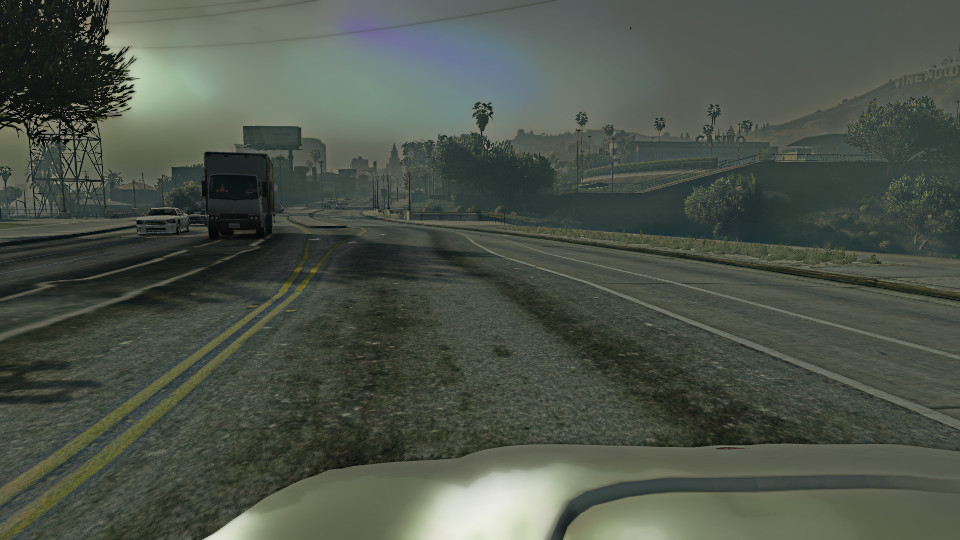} \hspace{5.5em}
  \includegraphics[width=.37\linewidth]{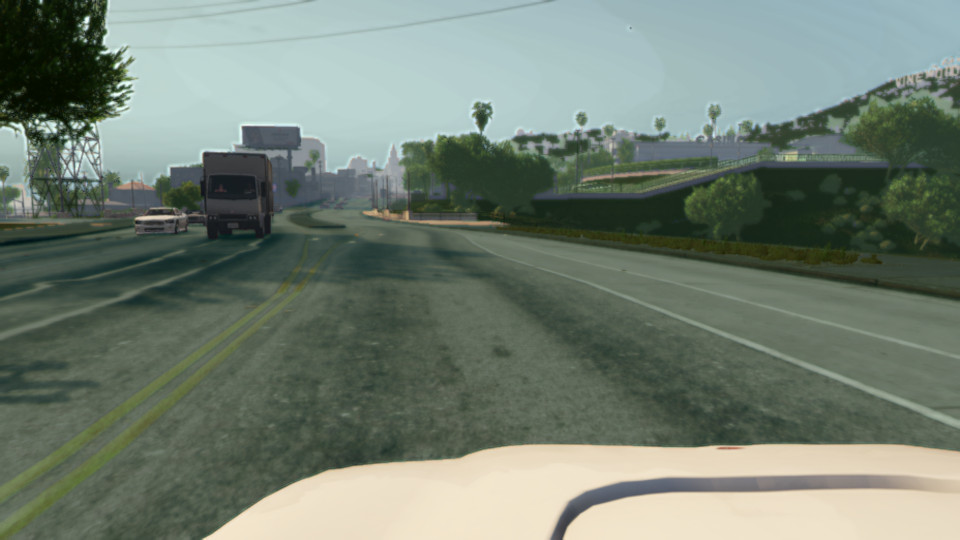} \\
  \begin{tabular}{p{0.45\textwidth}p{0.45\textwidth}}
  \centering CDT\cite{PITIE2007123} & \centering WCT2\cite{Yoo_2019_ICCV}
  \end{tabular}
  \includegraphics[width=.37\linewidth]{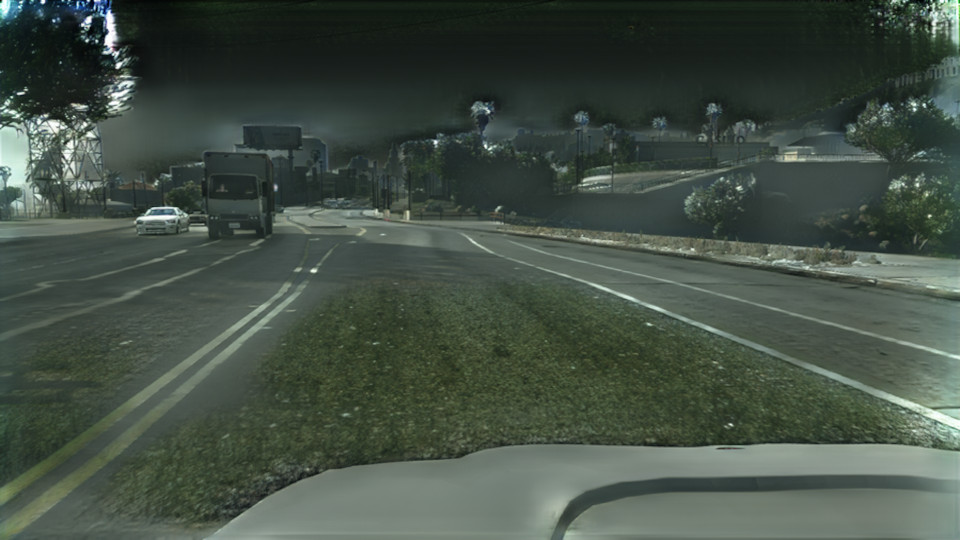} \hspace{5.5em}
  \includegraphics[width=.37\linewidth]{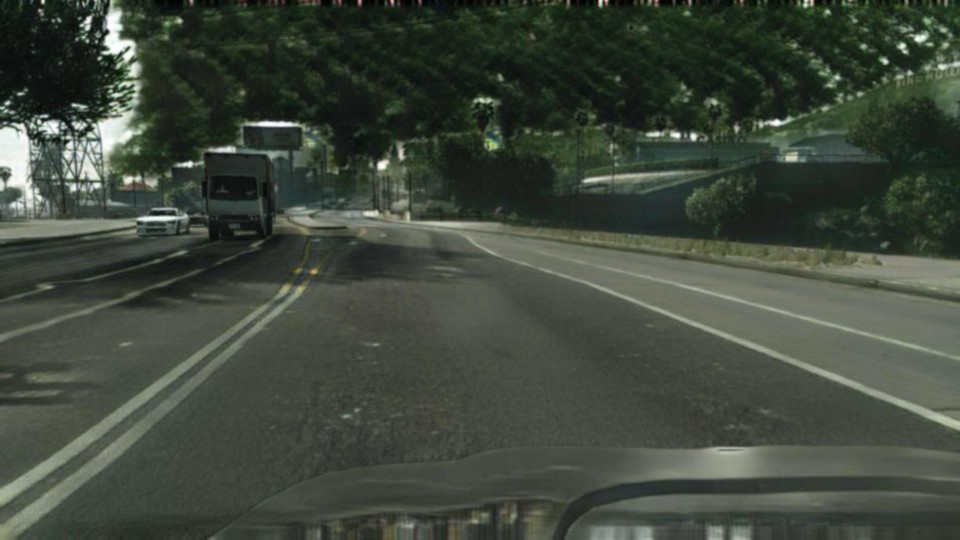} \\
  \begin{tabular}{p{0.45\textwidth}p{0.45\textwidth}}
  \centering TSIT\cite{jiang2020tsit} & \centering MUNIT\cite{huang2018munit}
  \end{tabular}
  \includegraphics[width=.37\linewidth]{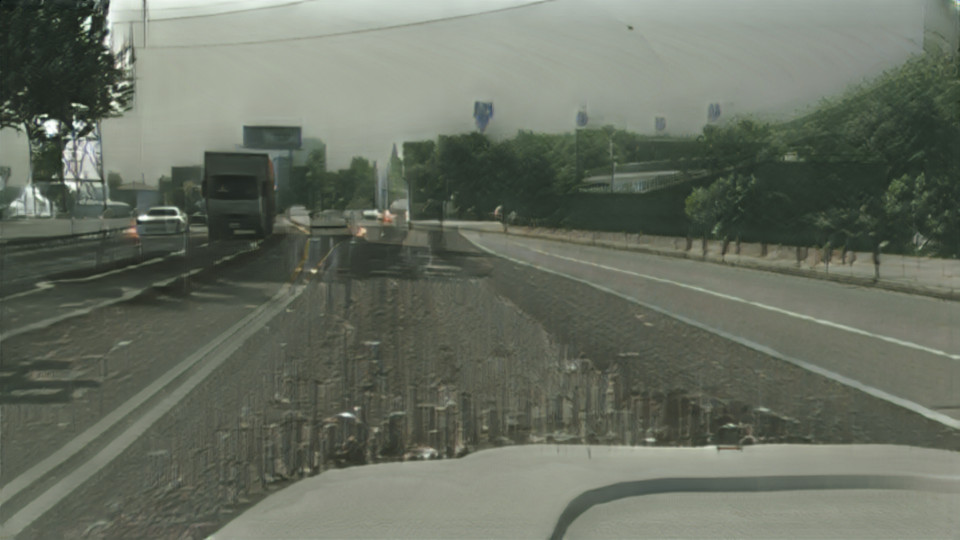} \hspace{5.5em}
  \includegraphics[width=.37\linewidth]{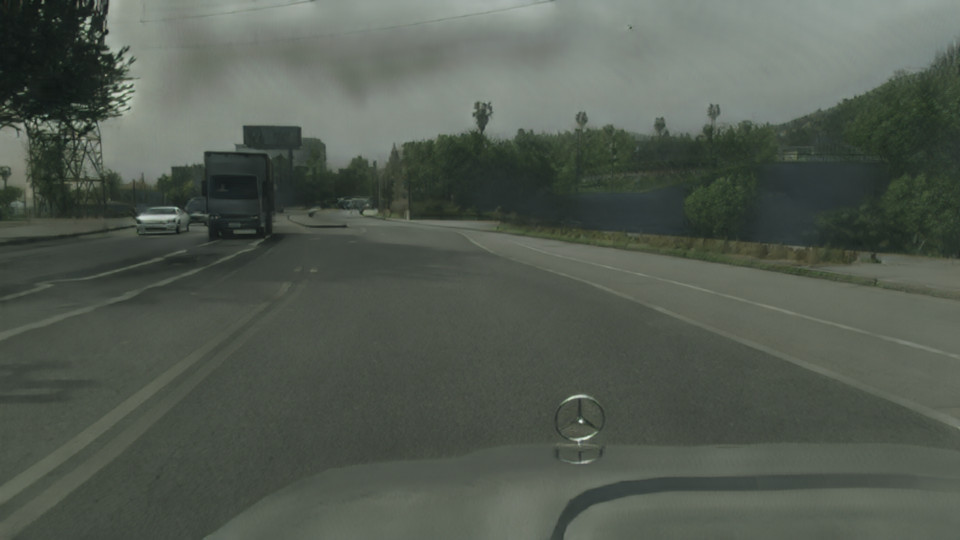} \\
  \begin{tabular}{p{0.45\textwidth}p{0.45\textwidth}}
  \centering CUT\cite{park2020cut} & \centering CyCADA\cite{Hoffman_cycada2017}
  \end{tabular}
  \includegraphics[width=.37\linewidth]{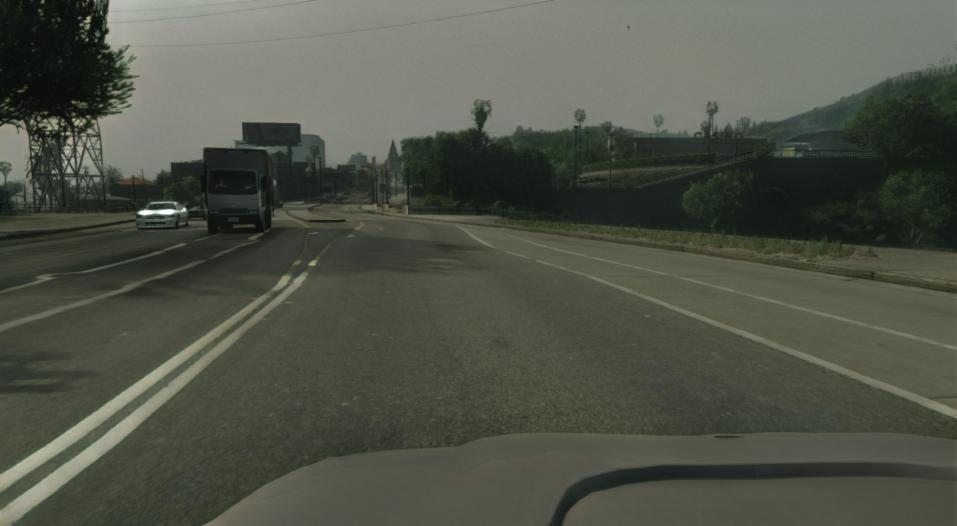} \hspace{5.5em}
  \includegraphics[width=.37\linewidth]{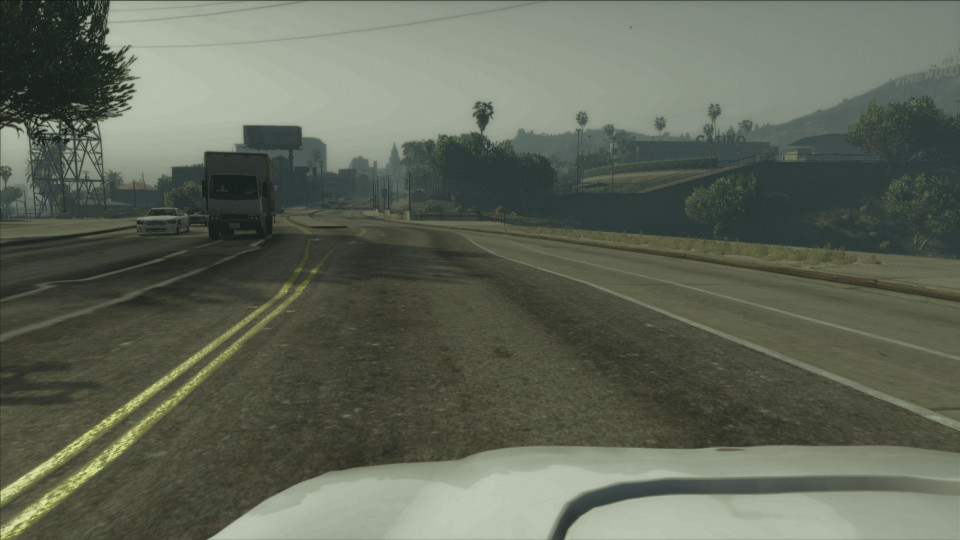} \\
  \begin{tabular}{p{0.45\textwidth}p{0.45\textwidth}}
  \centering EPE\cite{Richter_2021} & \centering Ours
  \end{tabular}  
  \caption{\label{fig:methods_comparison}%
            Comparison between our pipeline and other methods. Colour Transfer, CDT and WCT2
            fail to produce optimal results as they depend on favorable target images. TSIT, MUNIT,
            CUT and CyCADA hallucinate the content of the input frame by introducing undesired content such as 
            floating trees and high frequency artifacts.}
\end{figure*}

Outputs for the methods in this evaluation for the same input frame are given in Figure \ref{fig:methods_comparison} for comparison. Compared to methods relying on good alignment between the input frame and the style target, our method is capable of generalizing to a large set of input frames without producing unusual artifacts or leaving the input frame unchanged due to bad overlapping between the frames. With the exception of EPE, other unpaired image-to-image methods have a marked tendency to hallucinate unwanted content such as floating trees, car logos and high frequency artifacts. While EPE achieves impressive results in terms of realism enhancement, on certain scenes it can produce a notable halo effect around edges (Figure \ref{fig:halo_comparison}), possibly related to the segmentation maps used as an input by the generator. To demonstrate the temporal stability of the trained pipeline, our method has been applied to a series of video sequences from GTA 5 and from the \textit{GamingVideoSet} dataset \cite{Nabajeet2018}. 

Without the need for auxiliary metadata such as scene G-buffers or semantic maps, our method manages to produce structurally coherent and stable scenes, both spatially and temporally, while also enhancing realism.

\subsection{Ablation study}

To evaluate the effectiveness of each shader in enhancing the realism of a rasterized input, we conducted ablation experiments in which the shaders were introduced to the pipeline one at a time, and the pipeline retrained from scratch each time. The KID and FID metrics were used to quantitatively evaluate the contribution of each shader. The results are shown in Table \ref{tab:ablation}. 

The addition of shaders to the pipeline progressively reduces both metrics, demonstrating that each shader improves the similarity between the enhanced images and the target dataset, with the exception of a small drop in score when the noise shader is introduced. We suspect that the reason for this is that when noise is introduced into the images, information is lost, affecting the features extracted by the Inception model used in both metrics.

\subsection{Generalization to other Target Datasets and Input Content}

\begin{figure*}[tbp]
  \centering
  \begin{tabular}{p{0.48\textwidth}p{0.48\textwidth}}
  \centering\textbf{GTA (Rasterized Input)} & \centering\textbf{Cityscapes}
  \end{tabular}
  \includegraphics[width=.49\linewidth]{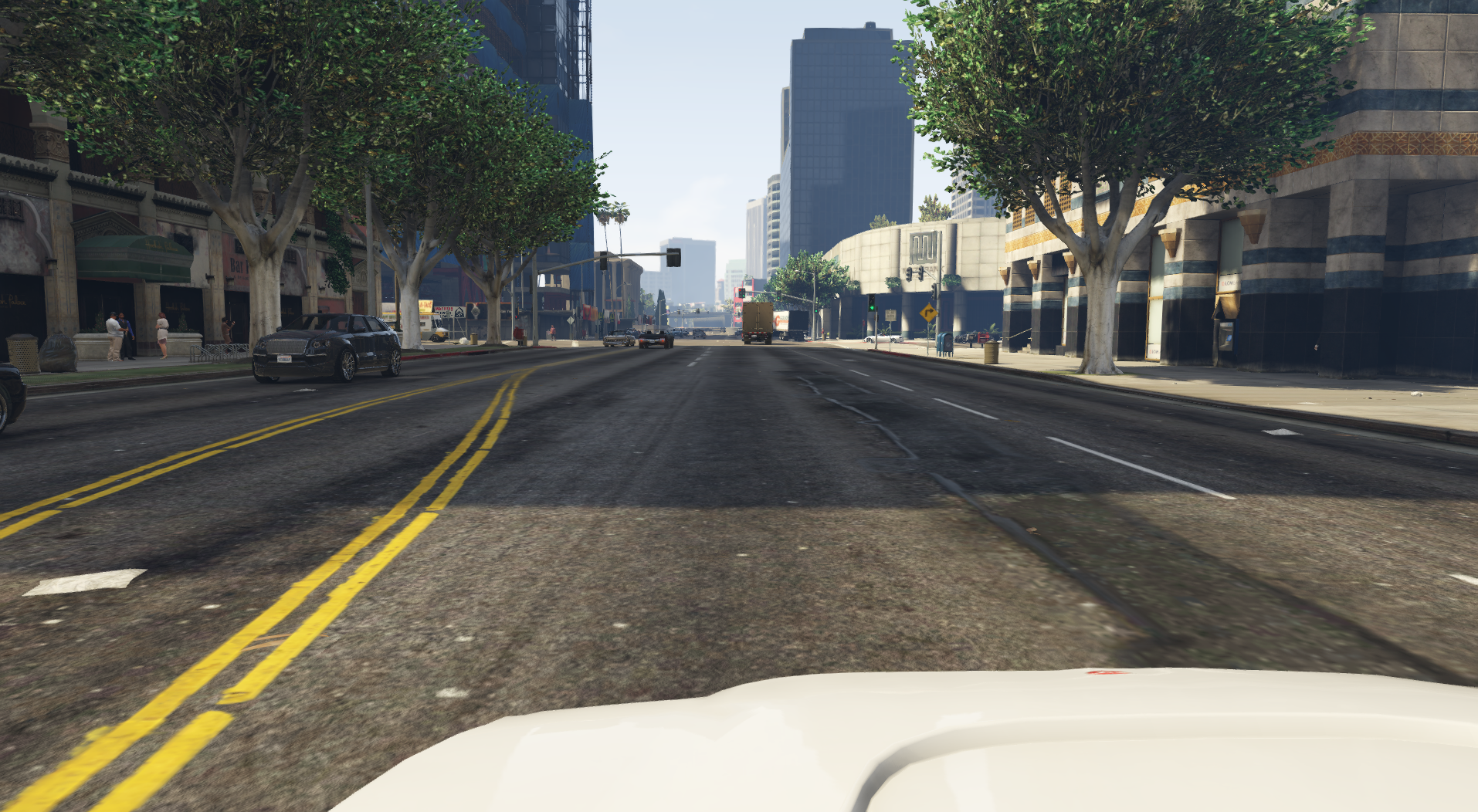}
  \hfill
  \includegraphics[width=.49\linewidth]{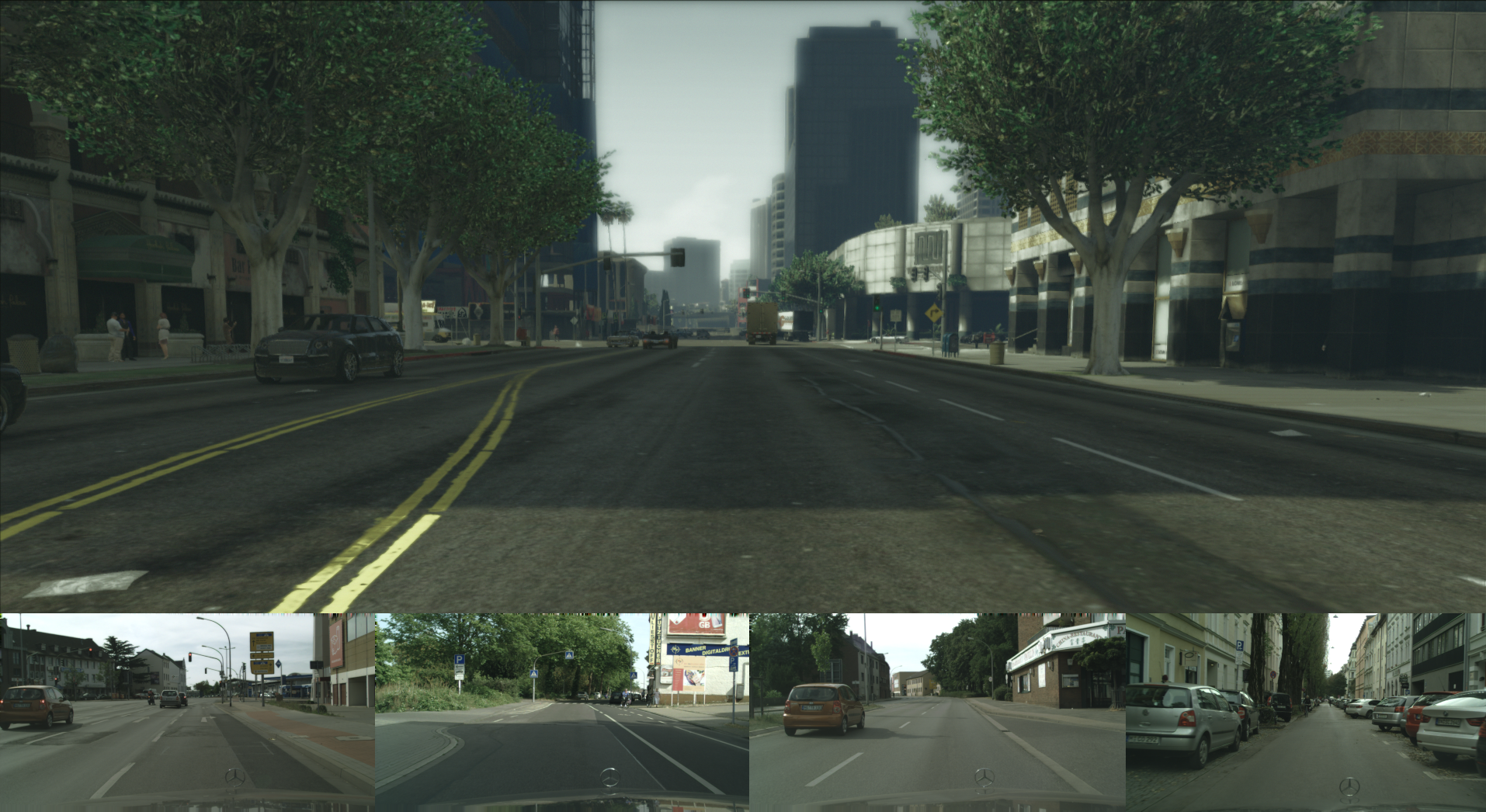} \\
  \begin{tabular}{p{0.48\textwidth}p{0.48\textwidth}}
  \centering\textbf{KITTI} & \centering\textbf{Mapillary Vistas}
  \end{tabular}
  \includegraphics[width=.49\linewidth]{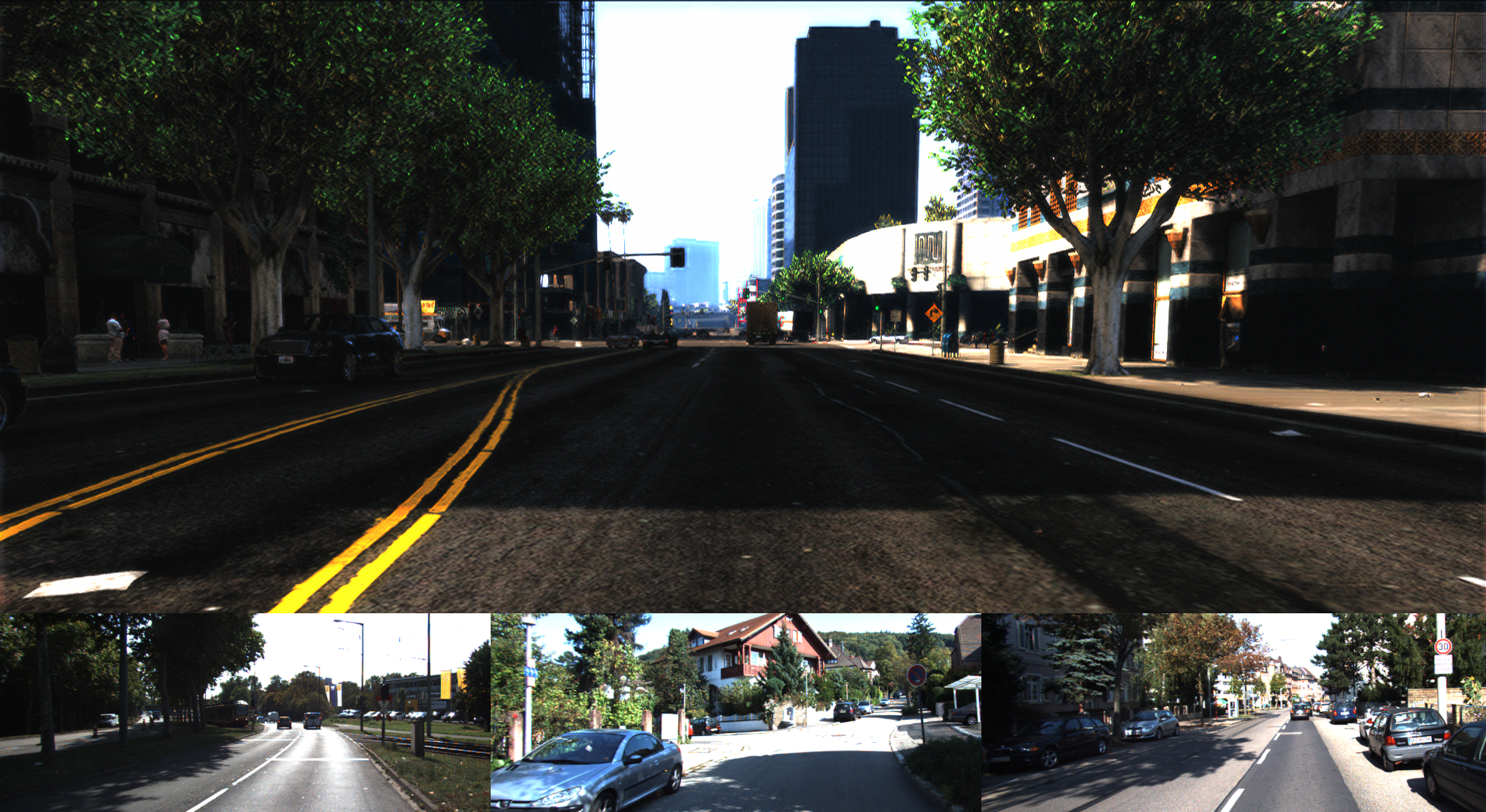}
  \hfill
  \includegraphics[width=.49\linewidth]{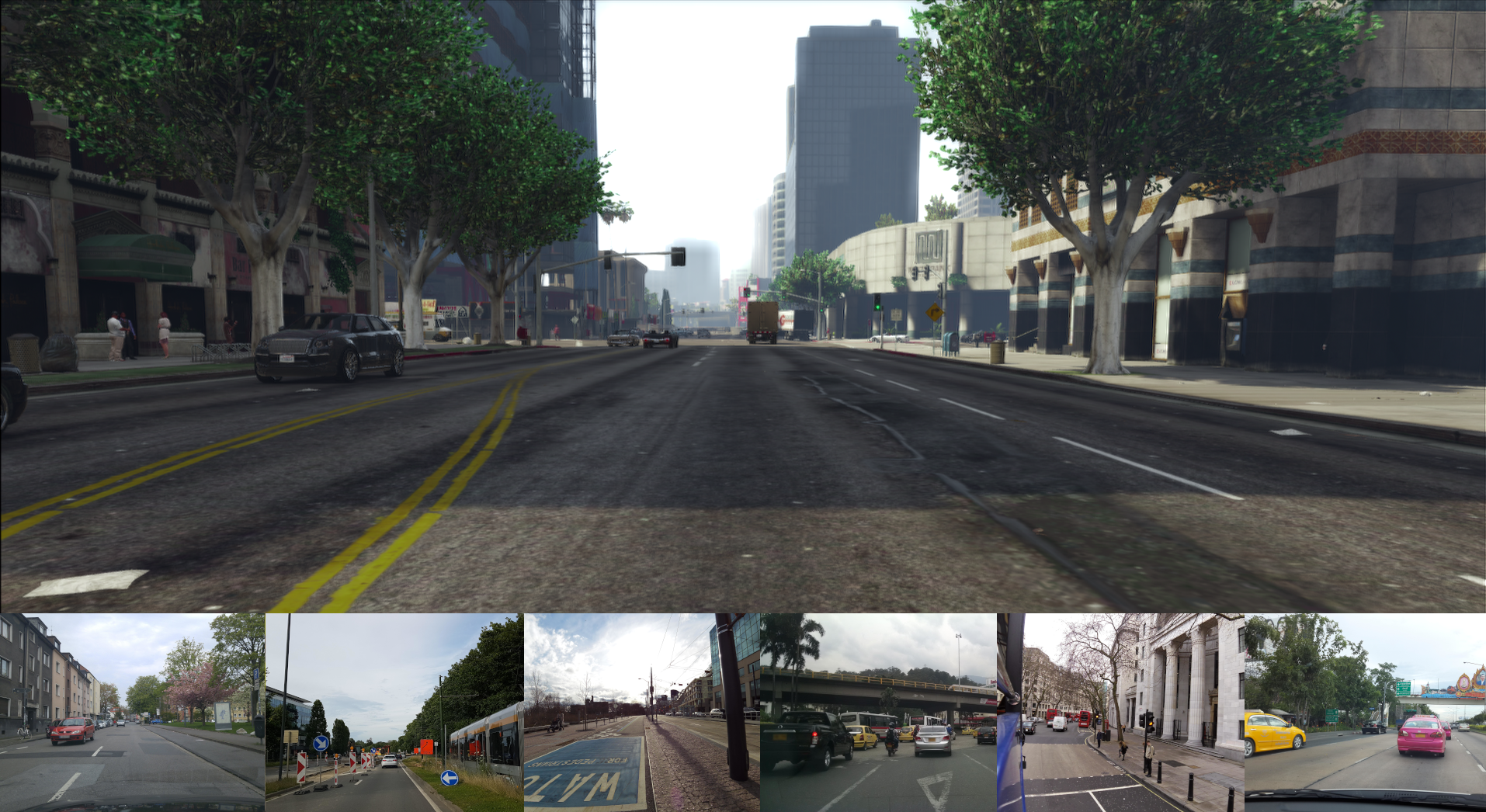}

  \caption{\label{fig:target_datasets_comparison}%
           Output of the proposed pipeline trained on multiple photographic datasets with different characteristic styles, showing that our proposed method
           generalizes well. Insets show images sampled from the respective target datasets.}
\end{figure*}
\begin{figure*}[h!] 
  \centering
  \setlength\tabcolsep{0pt}
  \begin{tabular}{c c c c}
	\textbf{Input Frame} & \textbf{Cityscapes} & \textbf{KITTI} & \textbf{Mapillary}\\
	\includegraphics[width=.25\linewidth]{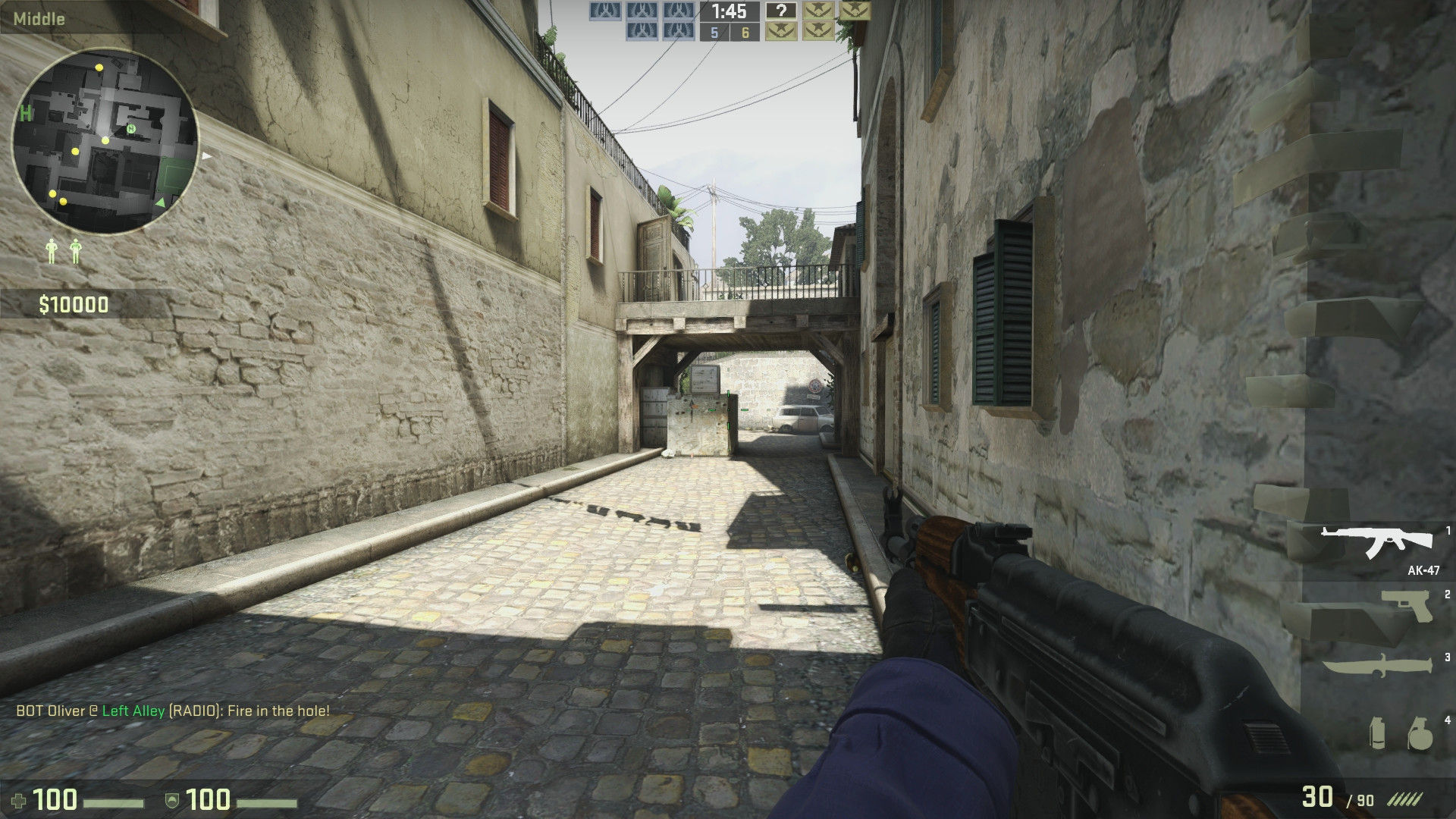} & \includegraphics[width=.25\linewidth]{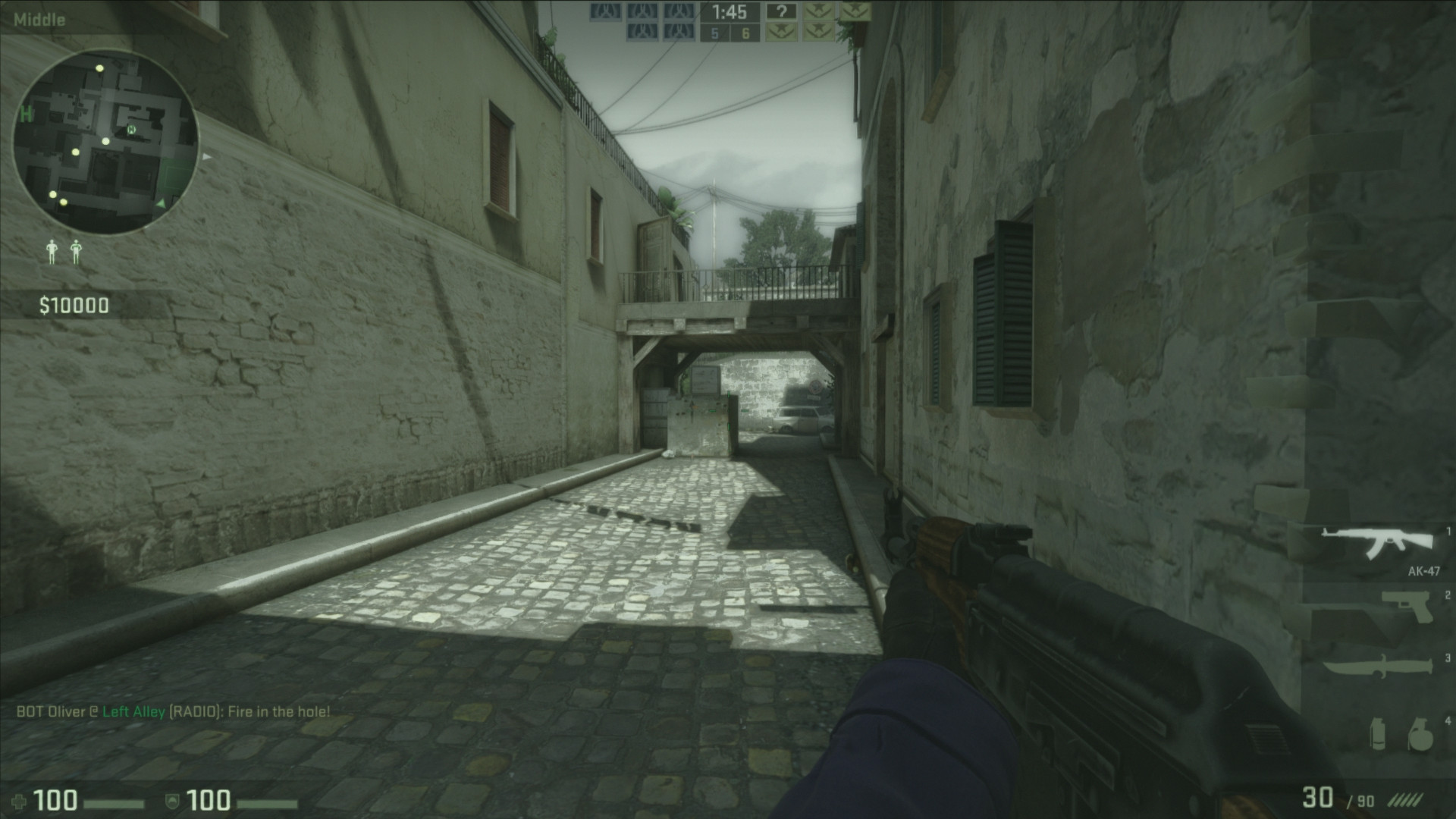} & \includegraphics[width=.25\linewidth]{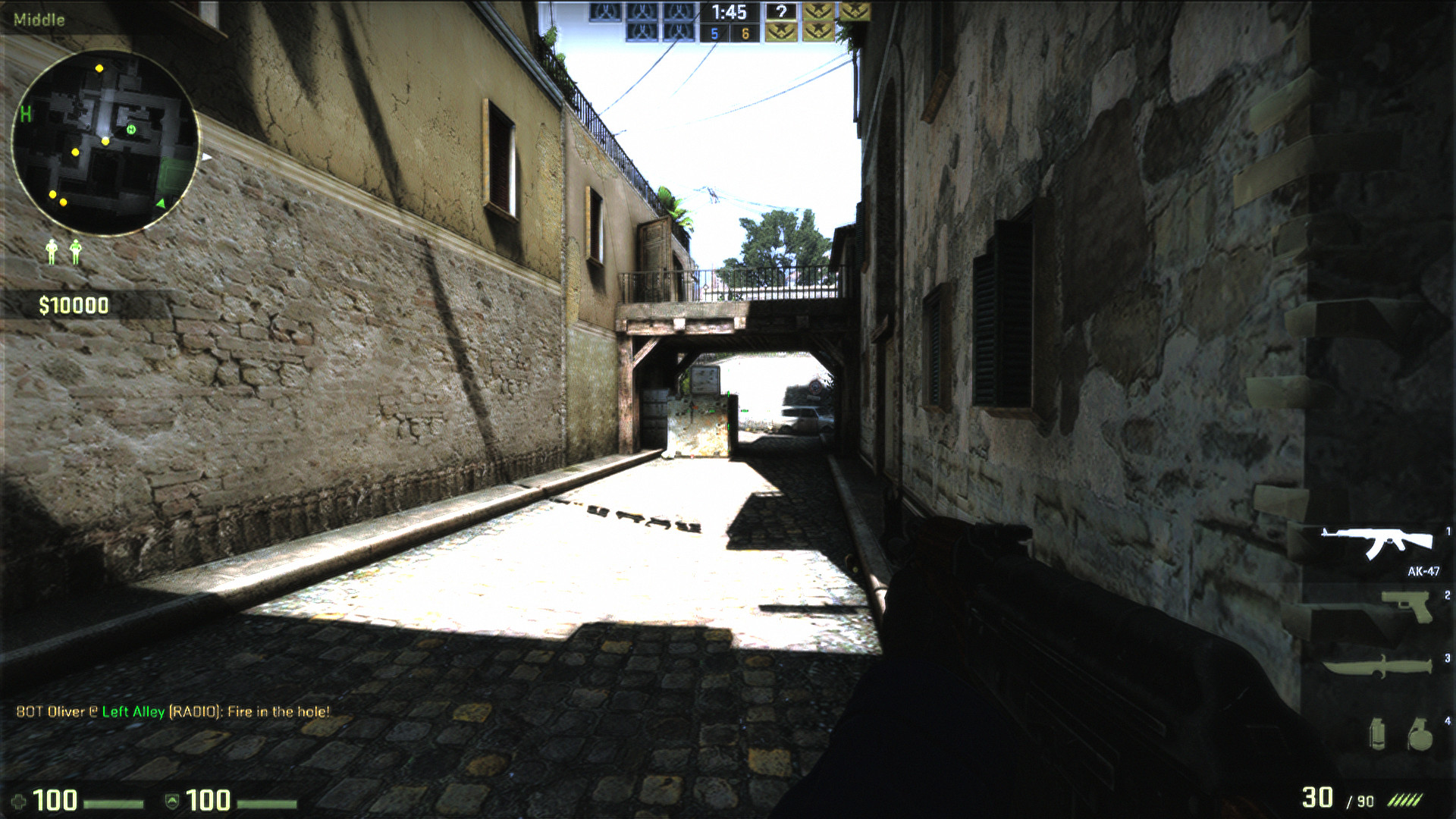} & \includegraphics[width=.25\linewidth]{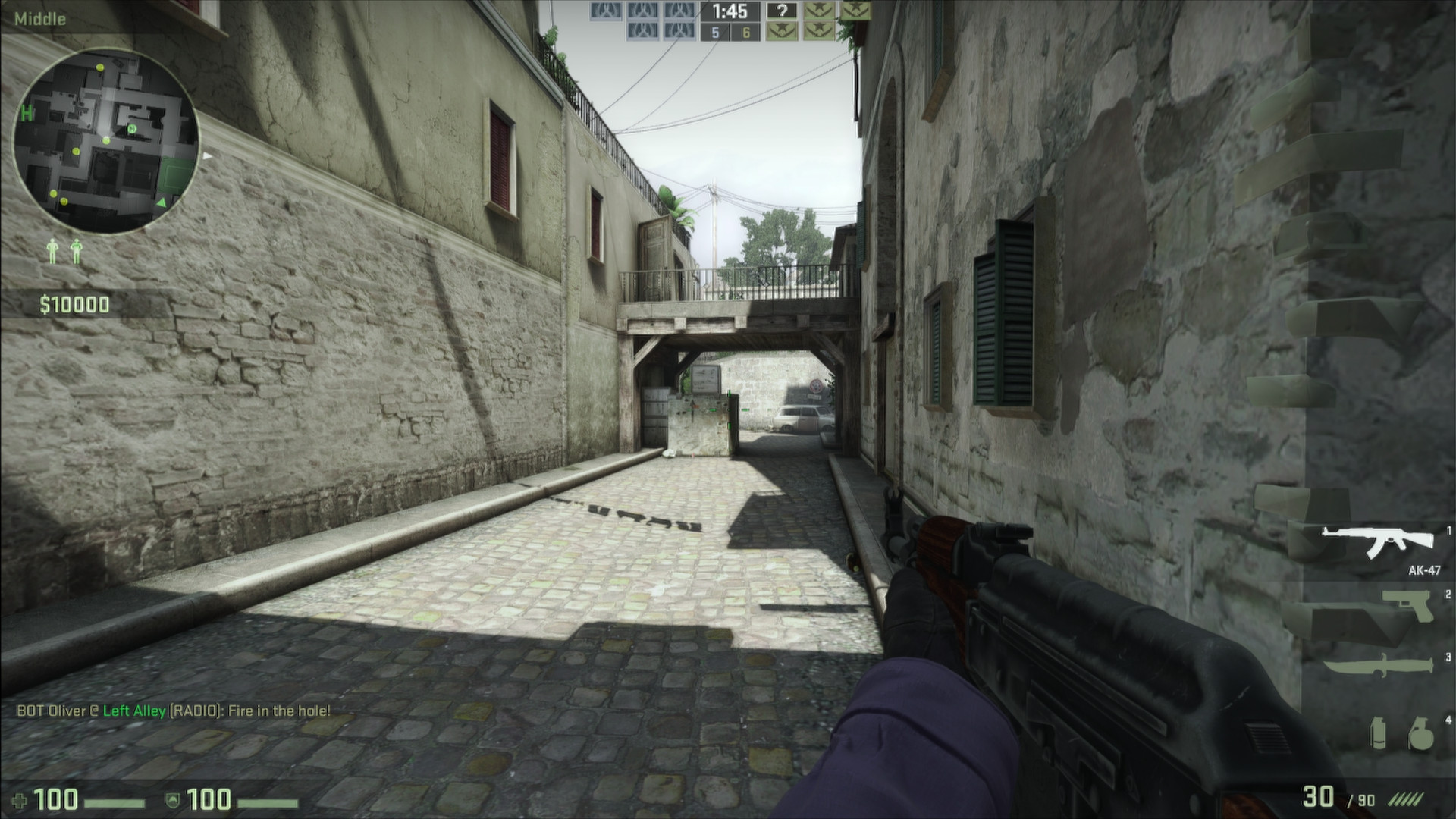}\\ 
	\includegraphics[width=.25\linewidth]{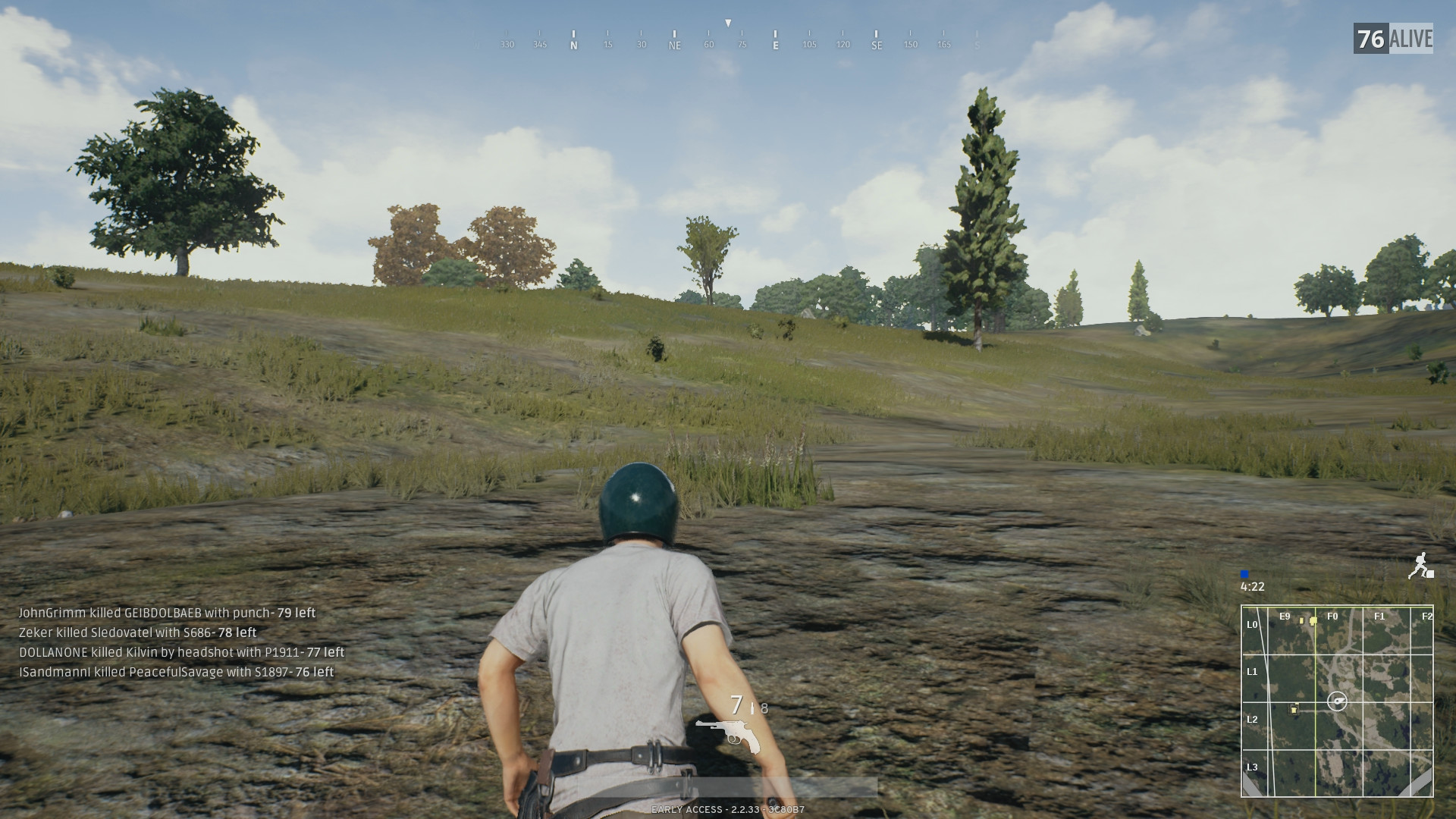} & \includegraphics[width=.25\linewidth]{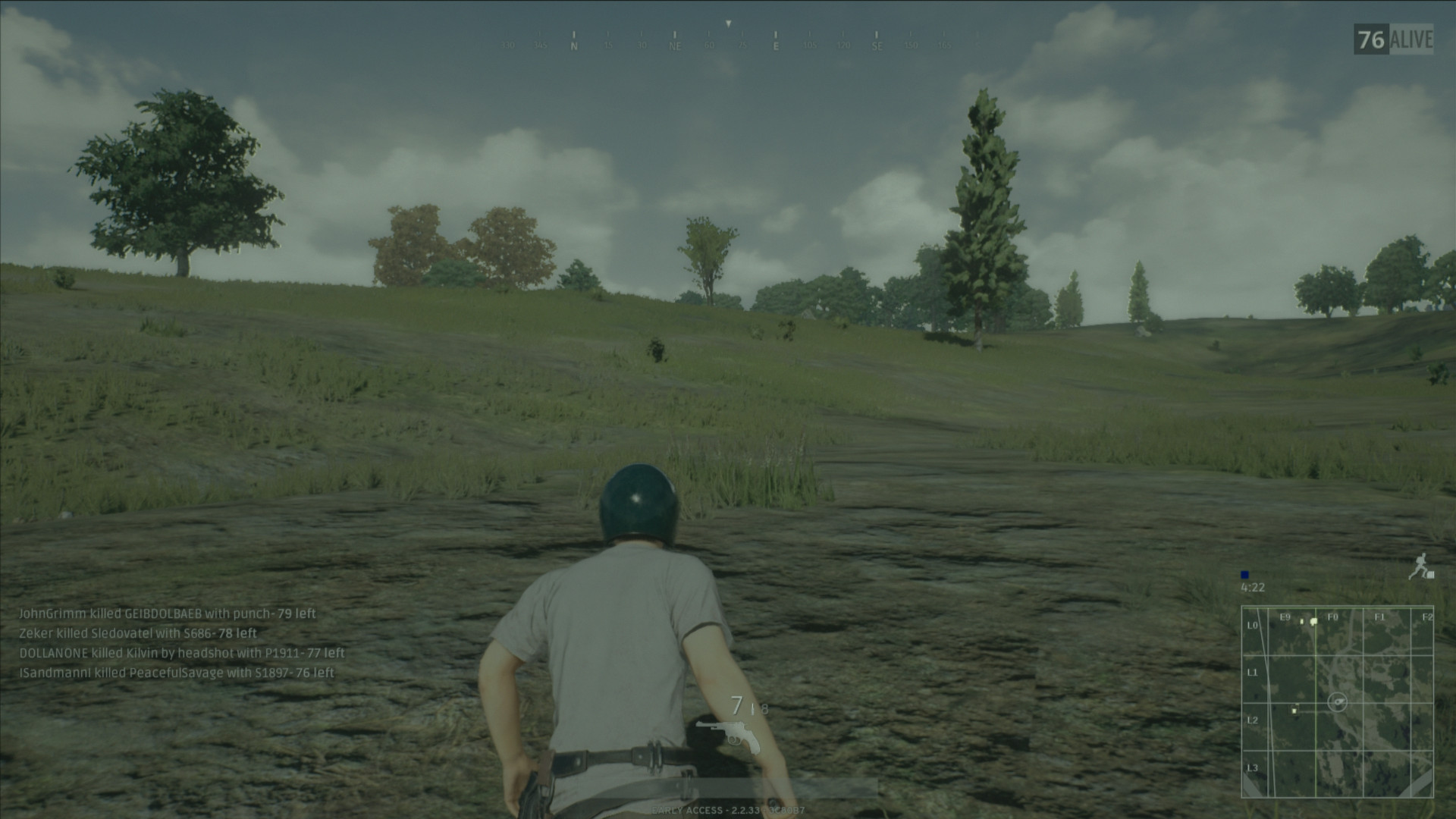} & \includegraphics[width=.25\linewidth]{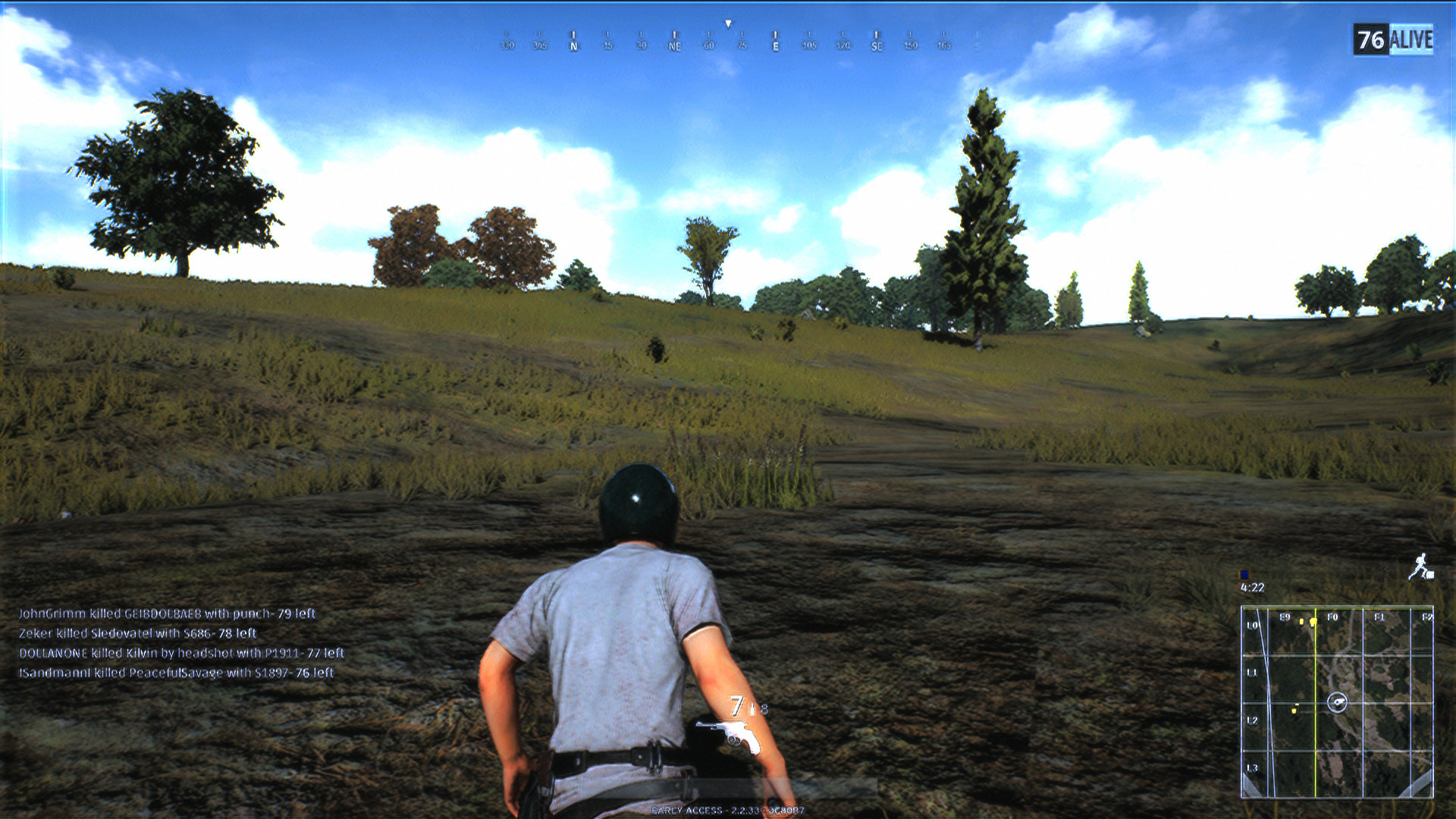} & \includegraphics[width=.25\linewidth]{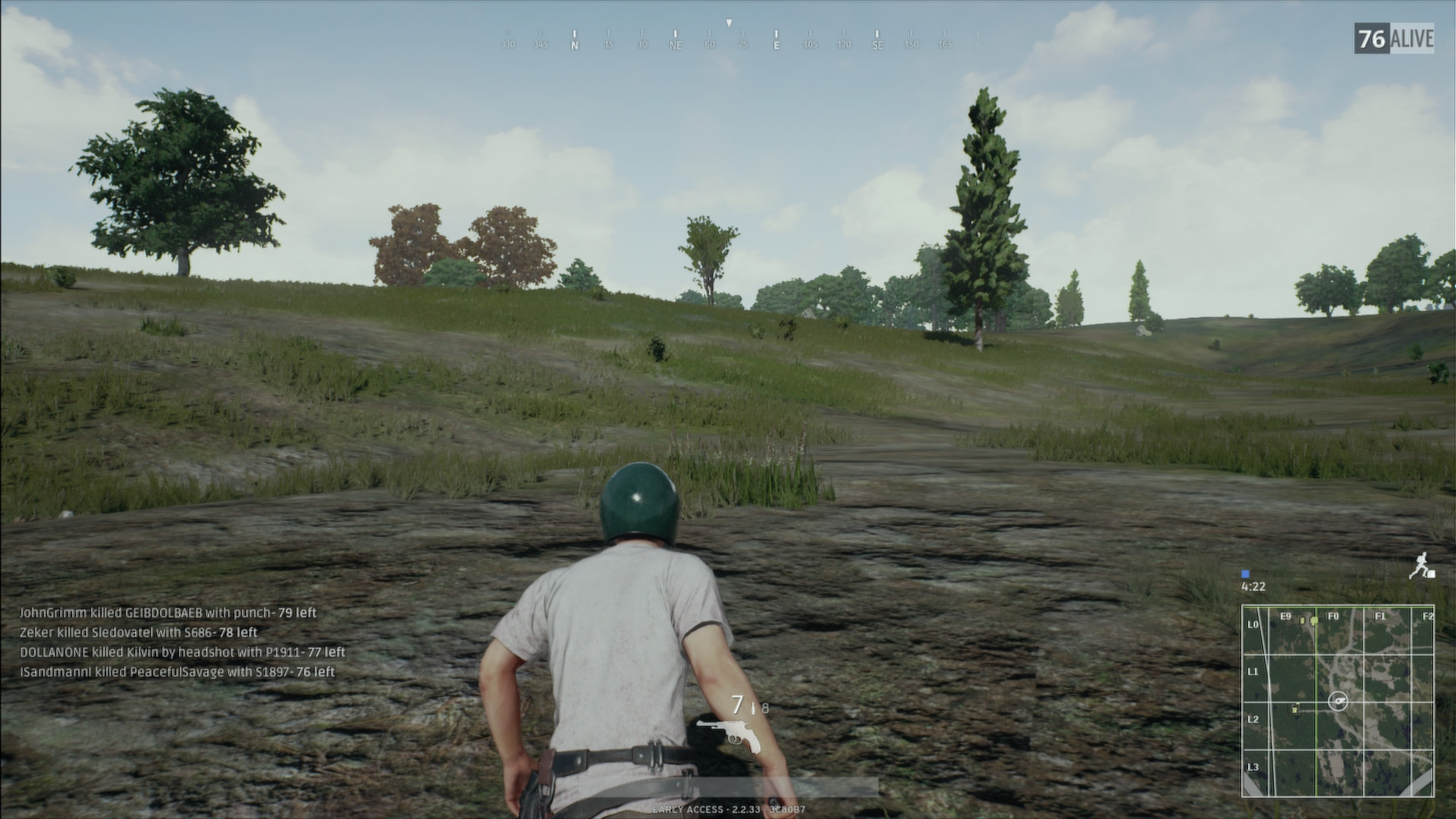}\\
	 \includegraphics[width=.25\linewidth]{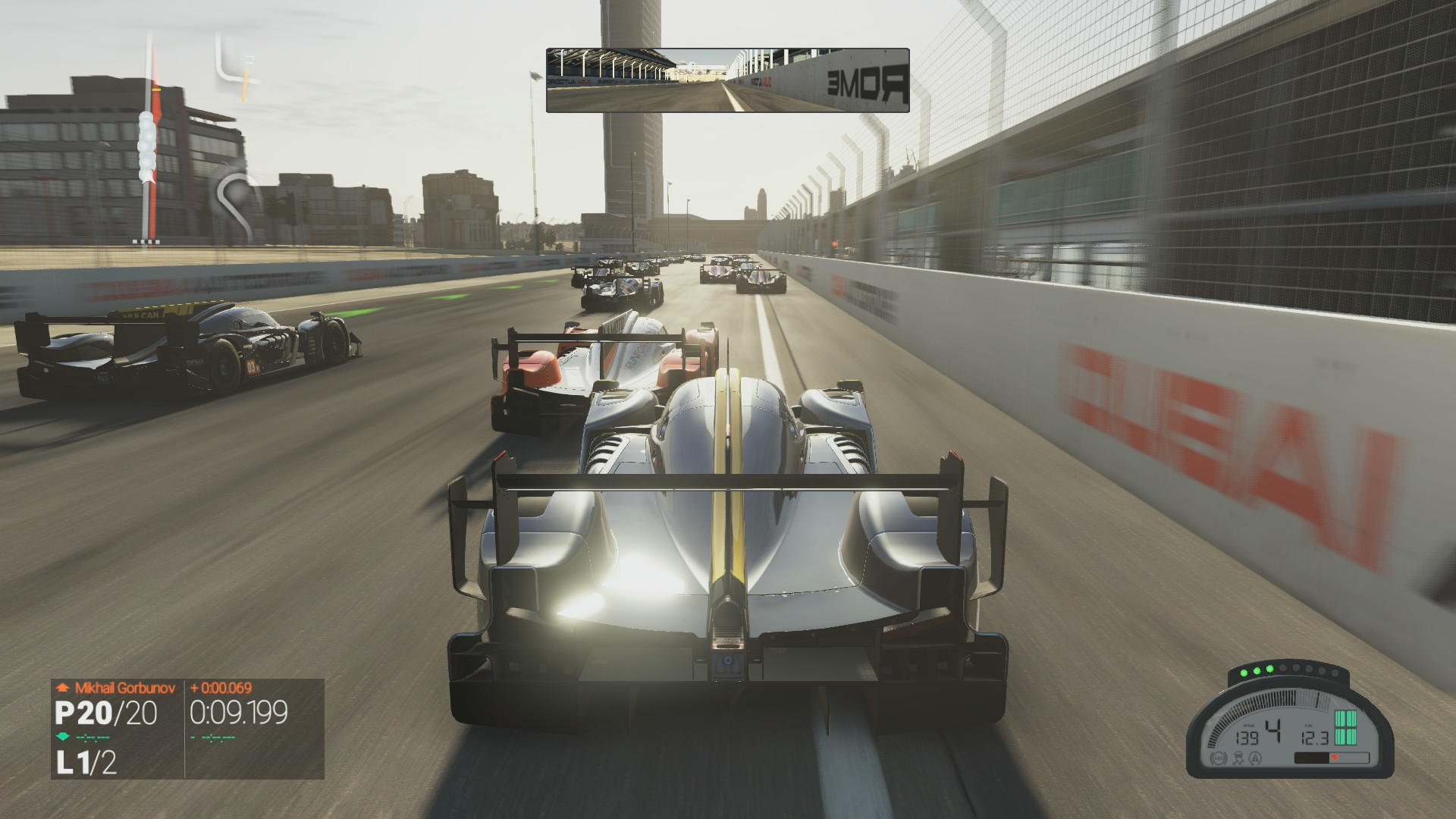} & \includegraphics[width=.25\linewidth]{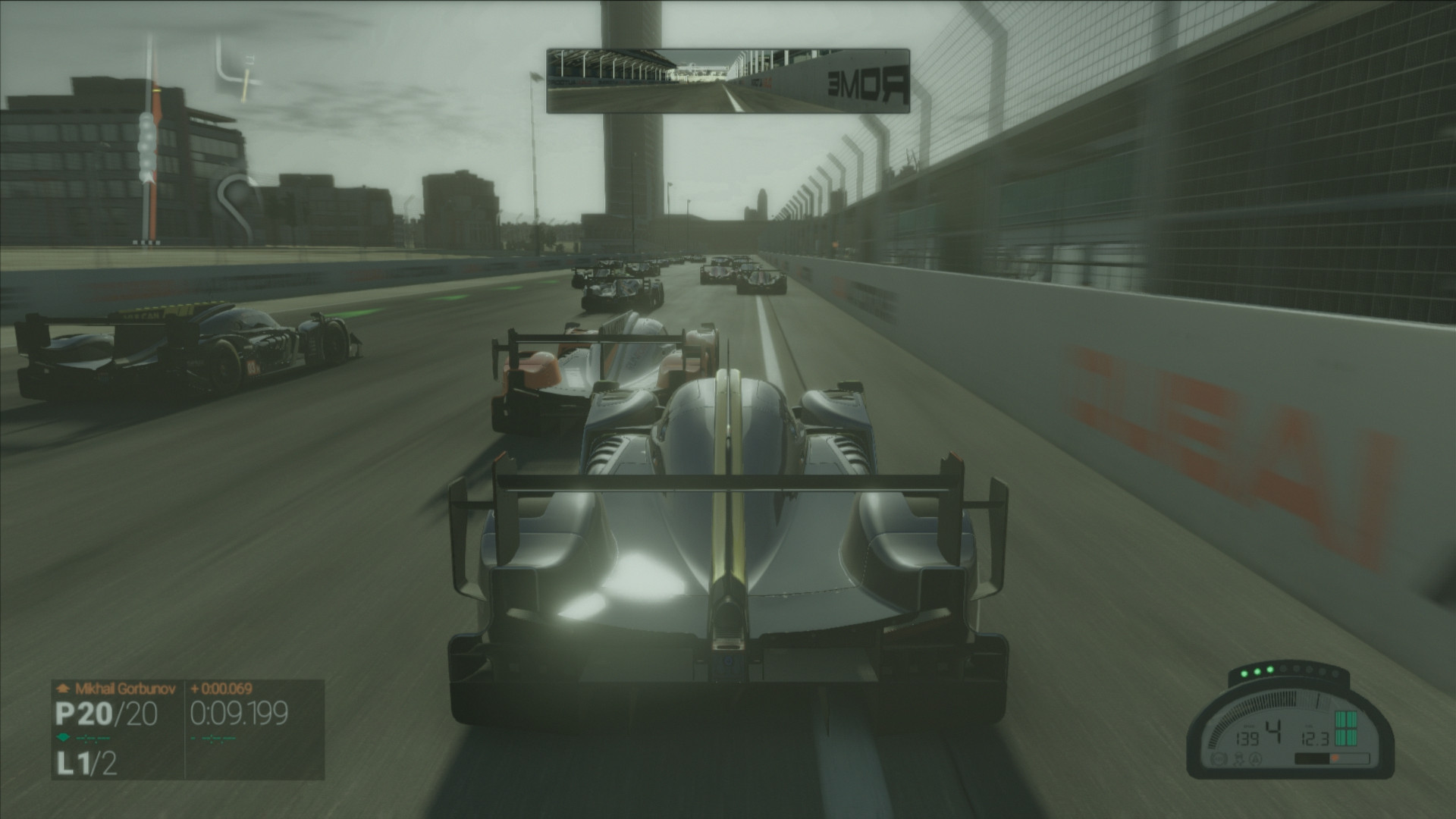} & \includegraphics[width=.25\linewidth]{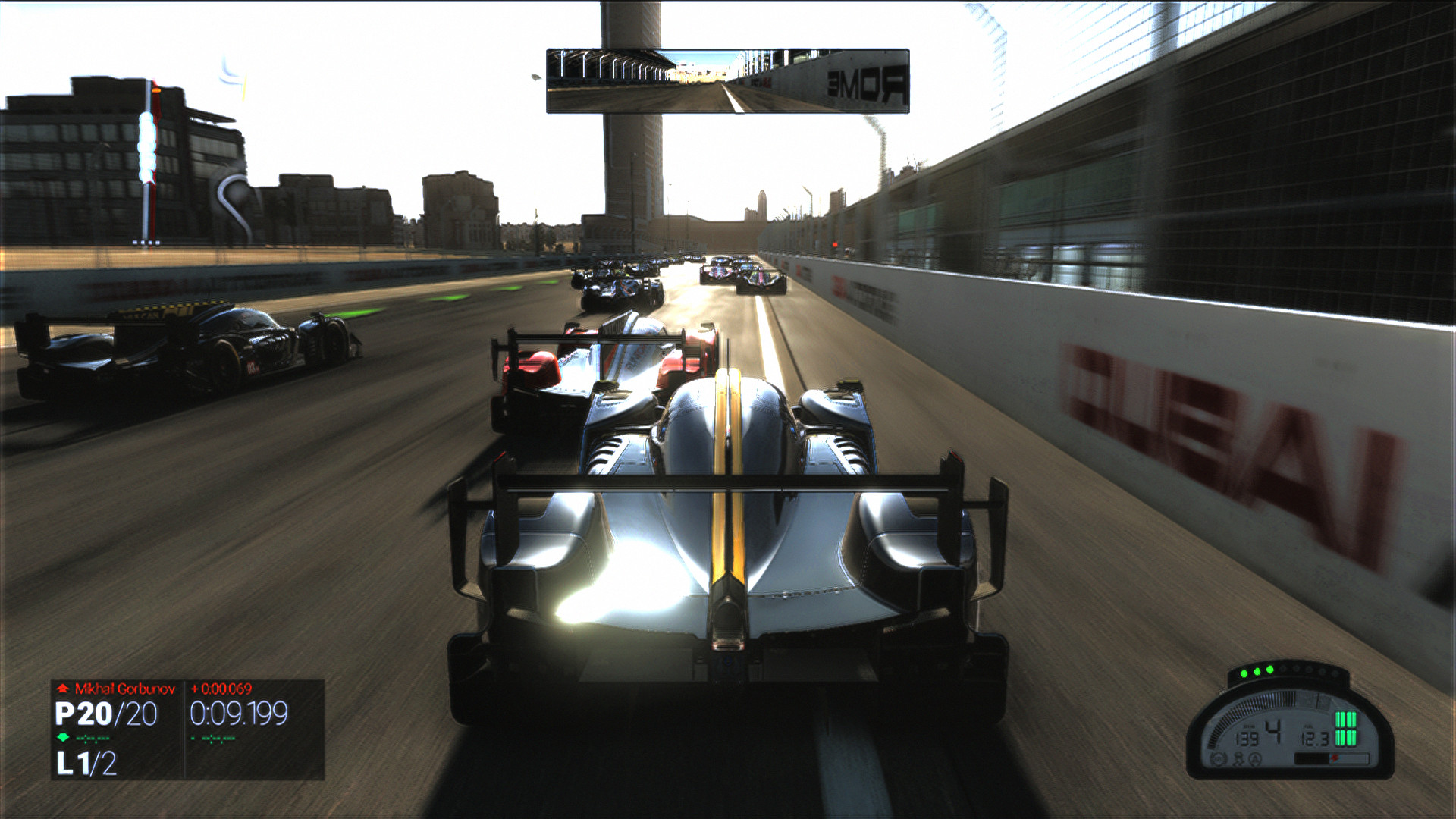} & \includegraphics[width=.25\linewidth]{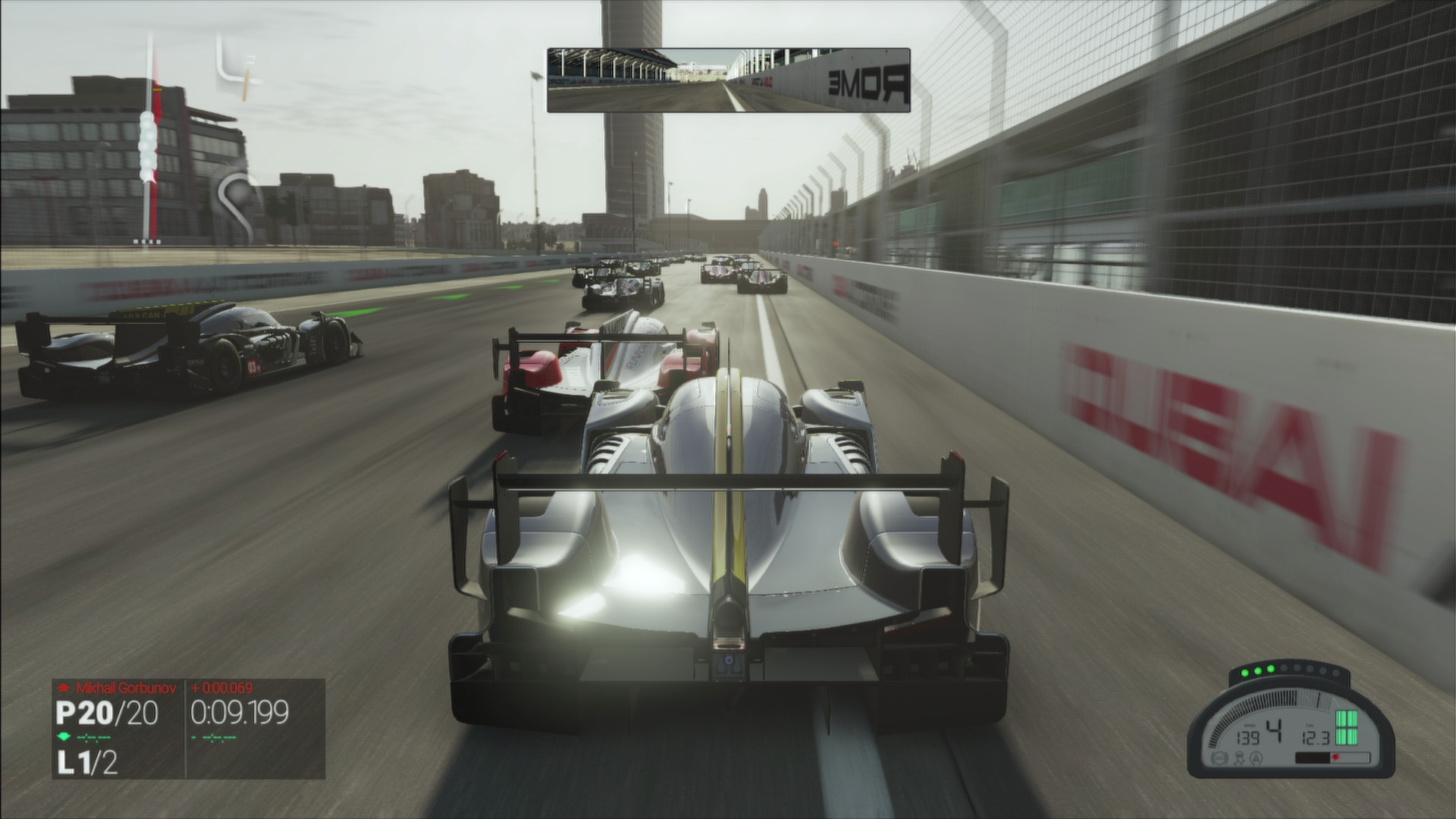}
  \end{tabular}
  \caption{\label{fig:input_content_comparison}%
           Output of the proposed trained pipeline when applied to different rasterised input content taken from \textit{GamingVideoSet} \cite{Nabajeet2018}, showing that our proposed method generalizes well to input content that has not been used during training.}
\end{figure*}

To demonstrate the generality of our approach, our Generative Adversarial Shaders were trained on Mapillary Vistas \cite{Mapillary2017} and KITTI \cite{Geiger2012CVPR} in addition to Cityscapes. In Figure \ref{fig:target_datasets_comparison}, we show some examples of the pipeline trained on KITTI and Mapillary Vistas.
The trained pipeline can also generalize well to different types of input content without needing to be re-trained or fine-tuned on different input datasets. To demonstrate this, we applied the trained pipeline on different frames taken from the \textit{GamingVideoSet} dataset \cite{Nabajeet2018}. The results are shown in Figure \ref{fig:input_content_comparison}.

\subsection{Limitations and Future Work}
\label{sub:limitation}
Our approach represents one end of what we expect is a continuum of methods ranging from real-time enhancement to high-quality, offline realism enhancement. Prioritising computational performance by using a series of handcrafted, learnable shaders instead of a large, undifferentiated neural network is a design choice that comes at the cost of limiting the range of effects our pipeline can replicate. For instance, unlike other models such as EPE \cite{Richter_2021}, our method cannot capture effects such as retexturing, introducing additional specularity or adding content in the input frame such as volumetric foliage. In future work these issues may be addressed by implementing "hybrid" techniques which combine the benefits of a bespoke shader pipeline with the learning capabilities of neural networks. To keep the pipeline lightweight and easily integrable, we avoided using additional input data such as depth maps and G-buffers. The use of this data may also be considered in future work as we believe it will also prove helpful when implementing such hybrid techniques.

Compared to other learning frameworks, adversarial training is known to be particularly unstable. In the case of our proposed method, such instabilities were exacerbated by the differences in learning capacities between the pipeline and the discriminator. In order to overcome these training limitations, adopting spectral normalization \cite{miyato2018spectral}, instance noise \cite{sonderby2014apparent},\cite{arjovsky2017towards} and removing edge artifacts from the pipeline output by cropping is necessary when training the model. Furthermore, in order to avoid undesirable interactions which can further reduce training stability, the parameters of the pipeline are designed to be mutually orthogonal. This imposes a design constraint which is also often a feature of more conventional, hand-tuned image processing pipelines.

Our approach is focused on camera modelling as an important component for realism enhancement of rasterized frames. Although this may restrict the applications of the proposed pipeline to style-transfer tasks involving the realism enhancement of 3D content, we expect that similar bespoke pipelines can be constructed for other domains by designing shaders that exploit relevant domain understanding.

\section{Conclusion}

We have introduced Generative Adversarial Shaders, combining specialized, performance-optimized post-processing shader pipelines with adversarial training. Using a bespoke pipeline that draws on an understanding of image formation yields several advantages. The inefficient black box of an undifferentiated neural network is replaced with a computationally efficient, easily-understood, interpretable and modular white box. An orders-of-magnitude reduction in the number of learnable parameters reduces the functional range of the model, which makes it impervious to deep learning related artifacts such as hallucination, and aids with temporal stability. The improved efficiency enables real-time realism enhancement, even on resource-constrained devices running a graphics rendering workload. The fact that only SDR images are required for input also makes the proposed pipeline straightforward to integrate.

We have shown that camera modelling accounts for a significant part of the realism enhancement problem. In future this can be combined with neural networks to (1) improve the overall efficiency of a deep learning based solution and (2) account for screen-space geometric effects (such as volumetric foliage \cite{Richter_2021}) that cannot easily be covered by such a shader pipeline.  We also expect that additional shaders can be added to the pipeline to model other image capture effects and further improve realism.

\section*{Acknowledgements}
We would like to thank our colleagues at Imagination Technologies for their help and support with this project. In particular, we would like to acknowledge Christian Burisch for sharing his thoughts on realism enhancement, and Kristof Beets for helpful discussions.
We thank the reviewers for providing us with constructive feedback on how we could improve our paper.
Lastly, we would like to thank Dr. Nabajeet Barman from Kingston University for granting us access to \textit{GamingVideoSet} at short notice.

\bibliographystyle{eg-alpha-doi} 
\bibliography{paper_bibliography}       

\end{document}